\documentclass[a4paper, 11pt]{article}

\usepackage{amsmath,amsfonts,amssymb,amsthm,commath,graphicx, epstopdf, lmodern, txfonts}
\usepackage{graphicx, caption, subcaption}
\usepackage[latin1]{inputenc}
\usepackage[english]{babel}
\usepackage[colorlinks=true]{hyperref}
\usepackage{xspace}

\usepackage{a4wide,epic}
\usepackage{dcolumn}
\usepackage{bm}
\usepackage{verbatim}
\usepackage{dictsym}
\usepackage{ulem}
\usepackage{amssymb}
\usepackage[latin1]{inputenc}
\usepackage{amsmath}
\usepackage{color}
\usepackage{epsfig}
\usepackage{booktabs, tabularx,multirow,array}

\newtheorem{assump}{Assumption}[section]
\newtheorem{proposition}{Proposition}[section]
\newtheorem{definition}{Definition}[section]
\newtheorem{lemma}{Lemma}[section]
\newtheorem{theorem}{Theorem}[section]
\newtheorem{cor}{Corollary}[section]
\newtheorem{remark}{Remark}

\newcommand{\Hi}{\mathcal{H}}
\newcommand{\Ei}{\mathcal{E}}
\newcommand{\Fi}{\mathcal{F}}

\newcommand{\pope}{\mathsf{p}}
\newcommand{\sope}{\mathsf{s}}
\newcommand{\qope}{\mathsf{q}}

\newcommand{\Y}{\mathcal{Y}}

\newcommand{\R}{\mathbb{R}}
\renewcommand{\qed}{\hfill $\Box$ \vspace{3mm}}

\newcommand{\Perm}{\mathfrak{P}}
\newcommand{\X}{\mathcal{X}}
\newcommand{\Si}{\mathcal{C}}

\begin{document}
\normalem	

\title{Extending robustness \& randomization from Consensus to Symmetrization Algorithms}
\author{Luca Mazzarella\thanks{Dipartimento di Ingegneria dell'Informazione,
Universit\`a di Padova, via Gradenigo 6/B, 35131 Padova, Italy ({\tt mazzarella@dei.unipd.it}).} \and 
Francesco Ticozzi\thanks{Dipartimento di Ingegneria dell'Informazione,
Universit\`a di Padova, via Gradenigo 6/B, 35131 Padova, Italy and Department of Physics and Astronomy,
Dartmouth College, 6127 Wilder Laboratory, Hanover, NH 03755, USA ({\tt ticozzi@dei.unipd.it}).} \and 
Alain Sarlette\thanks{INRIA Paris-Rocquencourt, QUANTIC project team, PSL Research University, France; and SYSTeMS, Ghent University, Technologiepark Zwijnaarde 914, 9052 Zwijnaarde(Gent), Belgium ({\tt alain.sarlette@inria.fr}).}}

\date{\today}

\maketitle 

\begin{abstract}
This work interprets and generalizes consensus-type algorithms as switching dynamics leading to symmetrization of some vector variables with respect to the actions of a finite group. We show how the symmetrization framework we develop covers applications as diverse as consensus on probability distributions (either classical or quantum), uniform random state generation, and open-loop disturbance rejection by quantum dynamical decoupling. Robust convergence results are explicitly provided in a group-theoretic formulation, both for deterministic and for randomized dynamics. This indicates a way to directly extend the robustness and randomization properties of consensus-type algorithms to more fields of application.
\end{abstract}


\section{Introduction}\label{sec:intro}

The investigation of randomized and robust algorithmic procedures has been a prominent development of applied mathematics and dynamical systems theory in the last decades \cite{Donoho2006cs,Motwani1995}. Among these, one of the most studied class of algorithms in the automatic control literature are those designed to reach {\em average consensus}~\cite{TsitsiklisThesis,ConsensusReview,Moreau2005,BoydGossip}: the most basic form entails a linear algorithm based on local iterations that reach agreement on a mean value among network nodes. 
Linear consensus, despite its simplicity, is used as a subtask for several distributed tasks like distributed estimation \cite{xiao2005scheme,dimakis2010gossip}, motion coordination \cite{leonard2007collective}, clock synchronization \cite{carli2011pi}, optimization \cite{ConvOptCons}, and of course load balancing; an example is presented later in the paper, while more applications are covered in \cite{ConsensusReview}. A universally appreciated feature of linear consensus is its robustness to parameter values and perfect behavior under time-varying network structure.

In the present paper, inspired by linear consensus, we present an abstract framework that produces linear procedures to solve a variety of (a priori apparently unrelated) \emph{symmetrization} problems with respect to the action of finite groups. The main practical contribution of this unified framework is a systematic approach to prove effectiveness and robustness of a whole class of switching algorithms where iterations are associated to convex combinations of linear actions of a finite group. Our results prove asymptotic convergence to symmetrization by focusing only on the way the iteration steps are selected, by studying a {\em lifted} dynamics. To this aim, only weak assumptions on the choice of possibly randomized actions applied at each iteration, and on the values of mixing parameters, are needed. Hence, the algorithms that converge in the proposed framework offer the same desirable features of linear consensus algorithms, including robustness and potential implementation in a randomized/unsupervised fashion. 

In the second half of the paper, we show that our linear symmetrization framework covers a diversified set of previously proposed algorithms, and can suggest some new ones for suitable problems: the only requirement is that they can be recast as a symmetrization problem. This naturally includes only a subset, comprising linear consensus, of distributed algorithms while many other relevant ones, like belief propagation \cite{pearl1986fusion,moallemi2006consensus}, distributed pagerank \cite{ishii2010distributed}, computations of other graph properties  \cite{wan2002distributed,awerbuch1987optimal}, or various algorithms for distributed data fusion in sensor networks do not directly belong to this class. On the other hand, our framework does directly cover a set of tasks and procedures which do not even involve a distributed network, but just have a common group-theoretic structure with consensus. For instance, we show how our framework unveils the robustness of {\em quantum Dynamical Decoupling} (DD)~\cite{viola-DD} protocols which are used for open-loop disturbance rejection in quantum control. Circuits generating random states, or gates for quantum information processing, can also be viewed in this light. In fact, symmetric and invariant states are ubiquitous in classical and quantum physics, and symmetry-breaking or -preserving dynamics are sought for a variety of tasks. In particular, in quantum control, symmetries are known to be associated to uncontrollable sectors of the space \cite{altafini-tutorial} or to subsystems that are protected from noise \cite{zanardi-symmetrizing,viola-generalnoise}; this seems to open the possibility for various future applications of our framework.

The paper is organized as follows. Section \ref{sec:gossip} outlines the main features of standard gossip consensus algorithms, that will serve as an inspirational and guiding example. Sections \ref{sec:main1} and \ref{sec:main2} develop our general framework, first relying on specific group actions and then moving to a general abstract framework. While the present paper mostly focuses on discrete-time dynamics, a natural continuous-time counterpart is introduced in Section \ref{ssec:main2:CT}, generalizing the idea first introduced in our paper \cite{OurMTNS2015} for a specific example. Section \ref{sec:convergence} proves convergence of general symmetrizing algorithms in deterministic and randomized settings. Finally, Section \ref{sec:apps} presents a diverse set of problems and existing algorithms that are covered by our general framework, {\em and for which we can claim the same robustness features of gossip-type algorithms.} In the appendix, an alternative proof of convergence of the lifted dynamics using relative entropy is proposed. 

{\em Notation:} Throughout the paper, we call a vector whose elements are nonnegative and sum to 1 a \emph{vector of convex weights}. We denote by $\vert S\vert$ the cardinality of a set $S$ (i.e.~the number of elements it contains).


\section{Guiding example: gossip iterations as randomized symmetrization}\label{sec:gossip}

Consensus-type problems are formalized by assigning local agents (subsystems) to vertices $1,2,...,m \in V$ of a graph and association a state $x_k(t)$ to each vertex $k \in V$. The possibility of an interaction between agent pairs  $(j,k)$ at time $t$ is modeled by the edges $E(t) \subset \{(j,k) : j,k \in V \}$ of the graph. We restrict ourselves to an undirected interaction graph, which identifies $(j,k)$ with $(k,j)$. The goal of \emph{consensus algorithms} is, by iterating interactions between subsystems starting from an arbitrary initial state $x_1(0),x_2(0),...,x_m(0)$, to reach a final state where $x_1=x_2=...=x_m$ at a value that reflects a given function of the initial values, e.g.~their mean. 

There are many variants of consensus algorithms, and here as an example we consider {\emph{linear gossip}~\cite{BoydGossip}, with $x_k$ belonging to $\R^n$ for $k=1,2,...,m$.} At each iteration, a single edge $(j,k)$ is selected from the set $E(t)$ of available edges at that time; the agents then update their state according to:
\begin{eqnarray}
\nonumber	x_j(t+1) & = & x_j(t) + \alpha(t) (x_k(t)-x_j(t))\\
\nonumber	x_k(t+1) & = & x_k(t) + \alpha(t) (x_j(t)-x_k(t))\\
\label{eq:classical-gossip}	x_\ell(t+1) & = & x_\ell(t) \quad \text{ for all } \ell\notin \{ j,k\} \, ,
\end{eqnarray}
where $\alpha(t) \in [\underline{\alpha},\overline{\alpha}] \subset (0,1)$. If $\alpha=1/2$, agents $j$ and $k$ move to the same point that is the average of their states. By iterating this rule, one hopes that all $x_j(t)$ asymptotically converge to the average of the $x_j(0)$.

The way in which the edges are selected over time leads to different evolutions for the whole system. We consider the following situations:
\begin{itemize}
\item \emph{Cyclic interaction:} at each time $t$ one link $(j(t),k(t))$ is selected deterministically by cycling through the elements of a time-invariant edge set $E$.
\item \emph{Random interaction:} at each time $t$ one link $(j(t),k(t))$ is selected at random, $(j(t),k(t))$ being a single-valued random variable onto the edge set $E(t)$.
\end{itemize}
A well-known result in the consensus literature is that gossip iterations --- both random and cyclic --- lead to consensus under sufficient graph connectivity assumptions. In addition, gossip evolutions preserve the total average  $\bar{x}=\tfrac{1}{m}\sum_{k=1}^m x_k$, so the state of each agent $k$ converges to $x_k=\bar{x}(0) = \bar{x}(t)$ for all $t$.
\begin{proposition}\label{prop:00}
\cite{BoydGossip,Moreau2005}~If there exists some $B>0$ (and $\delta>0$) such that the union of edges selected during $[t,t+B]$ form a connected graph for all $t$ (with probability $\geq \delta$), then iteration of \eqref{eq:classical-gossip} asymptotically leads to $x_k(t) = \bar{x}(0)$ for all $k$ (with probability $1$).
\end{proposition}

Summing up, gossip iterations thus perform a distributed asynchronous computation of the mean, in a robust way with respect to the network size and structure and to parameter $\alpha$, as long as the graph is not completely disconnected.\\

It is possible, however, to look at this gossip algorithm from another perspective. The evolution associated to \eqref{eq:classical-gossip} can be interpreted as a convex combination of two permutations, namely the trivial one (identity) and the transposition of the $j$ and $k$ state values:
\begin{eqnarray}
\nonumber	\left( x_j(t+1),\;x_k(t+1) \right) & = & (1\text{-}\alpha(t))\, \left( x_j(t),\;x_k(t) \right)\, + \, \alpha(t) \, \left( x_k(t),\;x_j(t) \right) \\
\label{eq:classical-gossipswap}	x_\ell(t+1) & = & x_\ell(t) \;\;\; \text{for all } \ell \notin \{j,k\}\, . \phantom{kk}
\end{eqnarray}
Let $\Perm$ denote the group of all permutations of the integers $1,2,...,m$ and for $\pi \in \Perm$ let $P_{\pi}$ be the unique linear operator such that $P_{\pi}\, (x_1,x_2,...,x_m) = (x_{\pi(1)},x_{\pi(2)},...,x_{\pi(m)})$ for any $x_1,x_2,...,x_m$.
It is easy to show that connectedness of a graph is equivalent to the property that the pairwise swaps associated to its edges generate the whole permutation group~\cite{Alggrpth}. By using linearity of \eqref{eq:classical-gossip} and basic group properties, it is also possible to show that the evolution up to time $t$ of the full state vector $x(t)=(x_1(t),\ldots,x_m(t))$ can always be written --- although maybe not uniquely --- as a convex combination of permutation operators on the initial states\footnote{This basic result will be proved in a more general setting later.}: 
$$ x(t)=\sum_{\pi \in \Perm} w_\pi(t) P_\pi\, x(0) \qquad \text{ with } \;\;\; w_\pi(t)\geq0,\;\; {\textstyle \sum_\pi} w_\pi(t)=1 \;\;\; \forall t \, .$$
Any map of this form obviously preserves the average $\bar x(t)$. The reformulation in terms of permutations defines consensus as being any state in the set
\begin{equation}\label{eq:defconsclass}
	\mathcal{C} = \{ x \in \X = \R^{mn} :\;  P_{\pi}\, x = x \; \text{ for all } \pi \in \Perm \} \, .
\end{equation}
Hence, consensus can be equivalently described as reaching a state that is invariant under (the action $P_\pi$ on $\X$ of) any element of the permutation group.

We call this \emph{symmetrization with respect to the permutation group.} In the next sections we develop a general framework to tackle symmetrization tasks by iterative, distributed algorithms. This allows for direct extension of the gossip consensus example to different state spaces, to networks that are more general than graphs, and to computational or control tasks not directly related to networks and consensus.


\section{Symmetrization from group actions}\label{sec:main1}
This section presents the key definitions and algorithmic elements of finite-group symmetrization on vector spaces. In particular, linear gossip can be seen as a particular case of this class of symmetrizing iterations. Further examples are developed in Section \ref{sec:apps}.

\subsection{Notation and Symmetrization Task}
Let $\mathcal{G}$ be a finite group, with number of elements $|\mathcal{G}|$. Let $\X$ be a vector space over a field $\mathbb{R}$ or $\mathbb{C}$, endowed with an inner product $\,\langle\,\, ,\,\,  \rangle: \X\times \X\longrightarrow \mathbb{C}\,$. 

We will consider a {\em linear action} of $\mathcal{G}$ on $\X,$ that is a linear map $a: \mathcal{G}\times \X\rightarrow \X$ such that 
$\quad a(hg,x) = a(h,a(g,x))\quad$ and $\quad a(\mathrm{e}_{\mathcal{G}},x)=x \quad$ for all $x\in \X$ and $g,h \in \mathcal{G}$, where $\mathrm{e}_{\mathcal{G}}$ is the identity of ${\mathcal{G}}.$ Note that this implies among others $a(g^{-1},a(g,x)) = x$. 
Although every linear action is associated to a representation\footnote{Given a group  ${\cal G}$, let  $\X$  be a vector space and let us denote the set of bijective linear transformations on $\X$ as $\mbox{GL}(\X)$. A representation of ${\cal G}$ is an homomorphism from ${\cal G}$ to $\mbox{GL}(\X)$, i.e. a map $\gamma:{\cal G}\longrightarrow \mbox{GL}(\X)$  such that $\gamma(gh)=\gamma(g)\gamma(h)\,\,\,\forall\,g,h\in{\cal G}$.} of $\mathcal{G}$ on $\X$, we maintain the action notation to make it directly applicable without re-parametrization, e.g.~when considering the conjugate action of the unitary group on quantum operators. From the inner product, we can define the adjoint of $a(g,\cdot)$ as the unique operator $a^{\dagger}(g,\cdot)$ that satisfies:
$\quad \langle y\, , \, a(g,x) \rangle=\langle a^{\dagger}(g,y)\, , \, x \rangle\quad \forall\,x,y\,\in \X\,.$

An element $\bar{x}\in \X$ is a fixed point of the action of $\mathcal{G}$ if 
\begin{equation}\label{def:symmetrization}
a(g,\bar{x})=\bar{x}\,\,\,\,\,\,\,\,\forall\,g \in \mathcal{G}.
\end{equation}
We denote the set of such fixed points as $\Si^{{\cal G}}\subseteq \X$. Since the action is linear, $\Si^{{\cal G}}$ is a vector space. Our main goal is the {\em symmetrization} of any initial condition $x\in \X$ with respect to the action of $\mathcal G,$ that is, construct an algorithm or a dynamical system that (asymptotically, with probability 1) drives any $x \in \X$ to some related $\bar x \in \Si^{{\cal G}}$. 

Consider any time-varying discrete-time dynamics $x(t+1) = \Ei_t(\,x(t)\,)$ on $\X$. We denote $\Ei_{t,0}(\cdot)$ the map associated to the evolution from time $0$ up to time $t$, such that $x(t) = \Ei_{t,0}(\,x(0)\,)$.  Let $\|\cdot\|$ be a norm associated to the inner product in $\X$.
\begin{definition}\label{def:goal0}
The algorithm associated to iterations $\{\Ei_t\}_{t\geq 0}$ attains asymptotic symmetrization if for all $x\in \X$ it holds:
\begin{equation}
\label{eq:11}
\lim_{t\rightarrow\infty}\|a(g,\Ei_{t,0}(x))-\Ei_{t,0}(x)\|=0\,\,\,\,\,\forall\,g\in\mathcal{G}.
\end{equation}
\end{definition}
We will also consider sequences of maps $\{ \Ei_t \}_{t\geq 0}$ that can be randomized; in this case, the above definition applies but convergence with probability one is understood. Note that for finite-dimensional $\X$, by linearity this implies uniform convergence. For infinite-dimensional $\X$, it would indicate a weak type of convergence.


\subsection{A Class of Algorithms}\label{ssec:algo1}

For a given group $\mathcal{G}$, vector space $\X$ and linear action $a:\mathcal{G}\times \X\rightarrow \X$, we will be interested in linear maps $\Fi$ of the form:
\begin{eqnarray}
	\label{eq:Eimap} \Fi(x) & = & \sum_{g \in \mathcal{G}} \, \sope_g \, a(g,x) \qquad \text{ with } \;\; {\textstyle \sum_{g \in \mathcal{G}}} \sope_g = 1 \; , \; \sope_g \geq 0 \; \forall g \, .
\end{eqnarray}
Such a map is completely specified by the choice of convex weights $\sope_g.$ From here on, we shall call a vector whose elements are nonnegative and sum to 1 a \emph{vector of convex weights}. We construct discrete-time dynamics on $\X$ by selecting at each time step $t$ a vector of convex weights $\sope(t)=(\sope_{g_1}(t),\sope_{g_2}(t),\ldots,\sope_{g_{|\mathcal{G}|}}(t)) \in  \R^{|{\cal G}|}$ and mapping $x(t)$ to $x(t+1)$ through the corresponding map of type $\Fi(x)$, i.e.
\begin{equation}\label{eq:xevol}
x(t+1) = \Ei_t(x(t)) := \sum_{g \in \mathcal{G}}\, \sope_{g}(t)\, a(g,x(t)) \, . 
\end{equation}
We assume that $\sope(t)$ is selected deterministically or randomly from some possibly infinite set ${\cal S}$. Typically any $\sope \in {\cal S}$ assigns nonzero weights only to a restricted set of $g \in \mathcal{G}$. From a dynamical systems perspective, we can interpret \eqref{eq:xevol} as a discrete-time {\em switching} system, whose generator is chosen at each time between a set of maps of the form \eqref{eq:Eimap}, according to the switching signal $\sope(t)$. The resulting $\Ei_{t,0}(\cdot)$ is also a convex combination of group actions, i.e.~of the form $\Fi(\cdot)$ given in \eqref{eq:Eimap}.

\begin{lemma}\label{lemma:convcomb} If the iterations have the form \eqref{eq:xevol}, then there exists a (possibly not unique) vector $\pope(t)=(\pope_{g_1}(t),\pope_{g_2}(t),\ldots,\pope_{g_{|\mathcal{G}|}}(t)) \in  \R^{|{\cal G}|}$ such that for any $t$ we can write:
\begin{equation}
\label{eq:totevol}
x(t)=\Ei_{t,0}(x(0))=\sum_{g\in\mathcal{G}}\pope_g(t)\, a(g,x(0))
\end{equation}
for any $x(0)$, with \newline
$\bullet$ at $t=0$, $\pope_{\mathrm{e}_{\mathcal{G}}}(0) = 1$ and $\pope_g(0) = 0$ for all $g\neq 0$ \newline
$\bullet$ for all $t$, $\sum_{g\in\mathcal{G}}\pope_g(t) = 1$ and $\pope_g(t)\geq 0 \; \forall g$.
\end{lemma} 
\proof
Proceed by inductive reasoning on $t$. For $t=1$, \eqref{eq:totevol} trivially holds because $\Ei_{1,0}(x)=\Ei_{0}(x)$ is given by \eqref{eq:xevol}. Now assume \eqref{eq:totevol} holds for some $t$. Then
\begin{eqnarray*}
\Ei_{t+1,0}(x)&=&\Ei_{t}\circ\Ei_{t,0}(x) \\
	\text{\scriptsize (def.$\Ei$)} &=&\sum_{h\in\mathcal{G}}\sope_h(t)a(h,\sum_{g\in\mathcal{G}}\pope_g(t)a(g,x))\\
	\text{\scriptsize (linearity)} &=&\sum_{h,g\in\mathcal{G}}\sope_h(t)\pope_g(t)\, a(h,a(g,x))\\
	\text{\scriptsize (def.action)} &=&\sum_{h,g\in\mathcal{G}}\sope_h(t)\pope_g(t)\, a(hg,x))\\
	\text{\scriptsize (var.change)} &=&\sum_{h,g'\in\mathcal{G}}\sope_h(t)\pope_{h^{-1}g'}(t)\, a(g',x))\\
									&=&\sum_{g'\in\mathcal{G}}\pope_{g'}(t+1)\, a(g',x)) \quad ,
\end{eqnarray*}
where we have defined $\pope_{g'}(t+1)=\;\sum_{h\in\mathcal{G}}\;\sope_h(t)\pope_{h^{-1}g'}(t).$ Noting that $g' \mapsto h^{-1} g'$ is a group automorphism such that $\sum_{g'\in\mathcal{G}}\, \pope_{h^{-1} g'}(t) = 1$ for each fixed $h$, one easily checks that $\pope(t+1)$ satisfies the requirements of a vector of convex weights. Hence the statement holds for $t+1$ and we get the conclusion by induction.\qed


\subsection{The symmetrizing map}\label{ssec:props1}
A general time-varying map might achieve symmetrization according to \eqref{eq:11} without ever converging to a fixed point. However, for dynamics of the form \eqref{eq:xevol} we have the following result.
\begin{proposition}\label{prop:3} An evolution defined by $\Ei_t$ of the form \eqref{eq:xevol} attains asymptotic symmetrization if and only if $\Ei_{t,0}(\cdot)$ converges to the fixed map
\begin{equation}\label{Fibar}
\bar{\Fi}(\cdot) \; := \; \frac{1}{|\mathcal{G}|}\sum_{g\in \mathcal{G}}a(g,\cdot) \,
\end{equation}
pointwise for all $x \in \X.$
\end{proposition}
\proof
Assume symmetrization is attained. Taking the (finite) sum of \eqref{eq:11} over all $g \in \mathcal{G}$, dividing by $\vert\mathcal{G}\vert$ and using the triangle inequality gives:
\begin{eqnarray}
\nonumber 0 &=& \lefteqn{\lim_{t\rightarrow+\infty} \left\Vert \tfrac{1}{\vert\mathcal{G}\vert}\, \sum_{g \in \mathcal{G}}\, a\left(\,g,\,\sum_{h\in \mathcal{G}} \pope_h(t) a(h,x)\,\right) \;-\; \Ei_{t,0}(x) \right\Vert}  \phantom{KKKKKKKK} \\
\nonumber \text{\scriptsize (linearity)} & = & \lim_{t\rightarrow+\infty} \left\Vert \tfrac{1}{\vert\mathcal{G}\vert}\, \sum_{g \in \mathcal{G},\,h \in \mathcal{G}} \pope_h(t)\, a\left(\,g,\, a(h,x)\,\right) \;-\; \Ei_{t,0}(x) \right\Vert \\
\nonumber \text{\scriptsize (def.action)} & = & \lim_{t\rightarrow+\infty} \left\Vert \tfrac{1}{\vert\mathcal{G}\vert}\, \sum_{g \in \mathcal{G},\,h \in \mathcal{G}} \pope_h(t)\, a(gh,x) \;-\; \Ei_{t,0}(x) \right\Vert \\
 \label{eq:limop2a}  \text{\scriptsize (var.change)} & = & \lim_{t\rightarrow+\infty} \left\Vert \tfrac{1}{\vert\mathcal{G}\vert}\, \sum_{g \in \mathcal{G},\,h' \in \mathcal{G}} \pope_{g^{-1}h'} (t)\, a(h',x) \;-\; \Ei_{t,0}(x) \right\Vert \\
 \label{eq:limop2b} \text{\scriptsize (see below)} & = & \lim_{t\rightarrow+\infty} \left\Vert \tfrac{1}{\vert\mathcal{G}\vert}\, \sum_{h' \in \mathcal{G}} \, a(h',x) \;-\; \Ei_{t,0}(x) \right\Vert
\end{eqnarray}
for all $x \in \X$, which would imply that $\Ei_{t,0}$ converges to $\bar{\Fi}$. To go from \eqref{eq:limop2a} to \eqref{eq:limop2b}, we sum on $g$ for each fixed $h'$: that yields $\sum_{g \in \mathcal{G}} \, \pope_{g^{-1}h'}(t) = \sum_{g' \in \mathcal{G}} \, \pope_{g'}(t) = 1$ for all $h'$, thanks to the facts that $g \mapsto g^{-1}$, and $g \mapsto gh$ (for fixed $h$), are group automorphisms.

For the converse:
Since both Definition \ref{def:goal0} and the present Proposition \ref{prop:3} concern pointwise convergence, we can as well assume a fixed $x$ and define  $b_g = a(g,x) \in \X$ for all $g \in \mathcal{G}$, that is a finite number of points in $\X$. Then any action just maps a $b_{g_1}$ to some other $b_{g_2}$, so the future evolution of the system can be restricted to the finite-dimensional linear subspace $\mathcal{B}$ of $\X$ spanned by the $b_g$. Then we have, since $a(h,\bar{\Fi}(\cdot)) = \bar{Fi}(\cdot)$ for all $h \in \mathcal{G}$ by definition, by linearity of the actions:
\begin{eqnarray*}
\Vert a(h,\Ei_{t,0}(x)) - \Ei_{t,0}(x) \Vert
& = & \Vert a(h,\,\Ei_{t,0}(x)-\bar{\Fi}(x)) + \bar{\Fi}(x) - \Ei_{t,0}(x) \Vert \\
& \leq & \Vert a(h,\,\Ei_{t,0}(x)-\bar{\Fi}(x)) \Vert +  \Vert \bar{\Fi}(x) - \Ei_{t,0}(x) \Vert \\
& \leq & (1+\bar{b}(x)) \, \Vert  \bar{\Fi}(x) - \Ei_{t,0}(x) \Vert
\end{eqnarray*}
where $\bar{b}(x)$ is an upper bound on the norm of the linear operator resulting from the restriction of $a(g,.)$ to the finite-dimensional vector space $\mathcal{B}$.\qed

The proof builds on the finite cardinality of $\mathcal{G}$ and remains valid if $\X$ is infinite-dimensional.
Notice however that if the actions associated to different $g \in \mathcal{G}$ are not all linearly independent, there will be more than one vector $\pope$ corresponding to the same map $\bar\Fi$ (see the next section).

\begin{lemma}\label{orthprog}
If there exists a group automorphism $g \mapsto h(g)$ such that 
\begin{equation}
\label{dagbi}
a^{\dagger}(g,\cdot)=a(h(g),\cdot) \;\;\; \forall g \in \mathcal{G}\; ,
\end{equation}
then $\bar\Fi$ is an orthogonal projection.
\end{lemma}
\proof
Eq.~\eqref{Fibar} readily yields that $\bar\Fi=\bar\Fi^2$ and that \eqref{dagbi} ensures $\bar\Fi=\bar\Fi^\dag$. 
\qed
Property \eqref{dagbi} holds e.g.~for any action that is a unitary representation of $\cal G$. Another advantage of a self-adjoint actions set is that it allows to easily determine a set of preserved quantities, depending only on the initial $x(0)$, as is the case for the mean in the gossip example.
\begin{lemma}\label{convquant}
If there exists a map (not necessarily an automorphism) $g \in \mathcal{G} \mapsto h(g) \in \mathcal{G}$ such that \eqref{dagbi} holds, then for any $\bar{z}\in \Si^{{\cal G}}$ we have
\begin{equation}
\langle \bar{z}, x(t) \rangle=\langle \bar{z}, x(0) \rangle \;\;\; \forall \, t \, .
\end{equation}\vspace{-2mm}
\end{lemma}
\proof
For any $t$ it holds that:
\begin{eqnarray}
\nonumber \langle \bar{z}, x(t) \rangle & = & \langle \bar{z}\,,\, \sum_{g\in \mathcal{G}}\pope_g(t)\,a(g,x_0) \rangle
\nonumber  \; = \;  \sum_{g\in \mathcal{G}}\pope_g(t)\, \langle a^{\dagger}(g,\bar{z})\, , \, x_0 \rangle  \\
\nonumber & = & \sum_{g\in \mathcal{G}}\pope_g(t)\, \langle a(h(g),\bar{z})\, ,\, x_0 \rangle
\nonumber \; = \; \sum_{g\in \mathcal{G}}\pope_g(t)\; \langle \bar{z},x_0 \rangle \, = \, \langle  \bar{z},x_0 \rangle\, .
\end{eqnarray}
\qed


\subsection{Example: linear gossip}\label{ssec:exmain1}

Consider the gossip algorithm described in Section \ref{sec:gossip}. To recast it in our framework, we choose $\X=\R^{mn}$ and ${\cal G}=\Perm$ the group of all permutations of $m$ elements. We can think of any $x \in \X$ as a column vector that stacks the $n$-dimensional state vectors of the $m$ subsystems. With the linear permutation operator $P_\pi$ defined Section \ref{sec:gossip}, the action of the group is simply $a(\pi,x)=P_\pi x.$ Notice that this action is self-adjoint. We have already established that consensus corresponds to the fixed points of this action, i.e.~$\mathcal{C} = \Si^{\Perm}$. From Proposition \ref{prop:3} and Lemma \ref{orthprog} (with the trivial automorphism $h(g) = g$), the map  $\; \bar{\Fi}=\frac{1}{m!}\,\sum_\pi\, P_\pi \;$ is the orthogonal projection onto the consensus set.

Next we turn to the evolution model. For linear gossip, the $m!$-dimensional vector $\sope(t)$ has only two nonzero entries at any time: $(1-\alpha(t))$ on the component corresponding to the group identity, and $\alpha(t)$ associated to swapping $j$ and $k$. If $\alpha$ and the graph with $\vert E \vert$ edges are constant, then $\sope(t)$ can switch between $\vert E \vert$ values. 
Let $P_{\mathrm{e}}$ and $P_{(j,k)}$ denote the linear operators $P_\pi$ that respectively implement the identity and the swapping of subsystems $j$ and $k$. These can be represented as $nm\times nm$ matrices: $P_{\mathrm{e}}=I_{nm}$, the identity, and $P_{(j,k)}=Q_{(j,k)}\otimes I_n$, the Kronecker product between the identity on $\mathbb{R}^n$ and $Q_{(j,k)}$ the $m\times m$ matrix that swaps the coordinates $j$ and $k$ of a vector of length $m$. Then the elementary evolution step associated to the selection of edge $(j,k)$ at time $t$ writes:
$$ x(t+1) \;=\; \sum_\pi\,\sope_\pi(t) \, a(\pi,x(t)) \;=\; (1-\alpha(t))\, P_{\mathrm{e}} x(t)+\alpha(t)\, P_{(j,k)}x(t)\, .$$

Finally, let us look at preserved quantities. Denoting $z_c$ the value on row $c$ of vector $z \in \X=\mathbb{R}^{mn}$, the set $\mathcal{C} = \Si^{\Perm}$ consists of all $z \in \X$ such that $z_{jn-d+1} = z_{kn-d+1}$ for all subsystems $j,k \in \{1,2,...,m\}$ and all components $d \in \{ 1,2,...,n\}$. This vector space is spanned in particular by the vectors $z^d \in X$, $d=1,2,...,n$, defined by:
$$
z^d_{jn-d+1} = 1/m \text{ for all } j\, , \, \text{other components 0}\,. 
$$
Hence by Lemma \ref{convquant}, we get as conserved quantities any linear functional of the form
$$
\langle \bar{z},x \rangle \;=\; \sum_{d=1}^{n} \, f_d \, \langle z^d,x \rangle \\
\;=\; \sum_{d=1}^{n} \, f_d \;\; \text{avg}(x)_d
$$
with arbitrary $f_1,f_2,...,f_{n} \in \mathbb{R}$, where $\text{avg}(x)_d$ denotes the average of the $d^{\rm th}$ component of the subsystem states.


\section{Action-independent dynamics}\label{sec:main2}

This section discusses {\em sufficient conditions} for obtaining symmetrization, that are {\em independent of the actions} but depend only on ${\cal G}$ and on the selected sequence of convex weights $\sope(t)$ at each step. These conditions are also \emph{necessary} if the particular actions associated to all elements of ${\cal G}$ are linearly independent. Since such actions exist for any finite group $\mathcal{G}$, the following conditions can be viewed as \emph{necessary and sufficient for obtaining symmetrization on all possible actions associated to a given group dynamics}\footnote{One representation with linearly independent elements is the {\em regular representation}: take $\X = \mathbb{R}^{|{\cal G}|}$, index the vectors of the canonical basis of $\X$ by $\{\, v(g) \in \X :  v(g)_h=\delta_{h,g} \,\,\,\forall\, g,h \in \mathcal{G}\,\}$ where $\delta_{g,h}$ is the Kronecker delta and define the linear action of $\mathcal{G}$ on $\X$ by $a(h,v(g)) = v(hg)$ for all $g,h \in \mathcal{G}$. To see that the actions associated to different $h \in \mathcal{G}$ are all linearly independent, it suffices to notice that $a(h,v(\mathrm{e}_{\mathcal{G}}))=v(h)$. This is essentially the representation used in \eqref{absevo}.}. In other words, we ensure asymptotic symmetrization for a general group-based algorithm in the form \eqref{eq:xevol} based only on the group properties and the selection rules for the convex vectors $\sope(t)$, for {\em any} underlying vector spaces and action. This frees us from the need to prove convergence for each specific application. Section \ref{sec:apps} provides a series of examples obtained by extending in this way the gossip-type algorithm.

More explicitly, Lemma \ref{lemma:convcomb} suggests that for studying the dynamics on $\X$ according to \eqref{eq:xevol}, it is sufficient to look at the evolution of the convex weights $\pope(t).$ The proof of the Lemma proposes the dynamics
\begin{equation}\label{prob1}
	\pope_g(t+1) \;=\; \sum_{h\in\mathcal{G}}\sope_{h}(t) \pope_{h^{-1} g}(t)
\end{equation}
for all $g \in \mathcal{G}$. If the group actions are linearly dependent, then several weights $\sope(t)$ or $\pope(t)$ can be associated to any map of the form $\Fi$ and clearly \eqref{prob1} is not the unique dynamics corresponding to \eqref{eq:xevol}. However, if we want to study \eqref{eq:xevol} by focusing on the group properties, and prove convergence in a way that is valid for {\em all possible actions associated to the group}, then \eqref{prob1} is the unique lift of \eqref{eq:xevol} that achieves this goal. In the current section we hence study the behavior of \eqref{prob1}.


Again, let us choose an ordering of ${\cal G}$ and consider $\pope(t)$, $\sope(t)$ as column vectors in $\R^{|{\cal G}|}$, i.e.~indices $g \in \mathcal{G}$ are identified with rows in the column vector. Then \eqref{prob1} becomes:
\begin{eqnarray}
\label{absevo}
\pope(t+1) & = & \left(\sum_{h\in \mathcal{G}} \sope_{h}(t) \Pi_{h}\right)\pope(t)\;=\;\tilde{M}(t)\pope(t) \;=\; \left(\prod_{i=0}^{t}\tilde{M}(i)\right)\pope(0) \, ,
\end{eqnarray}
where we define $\tilde{M}(t) = \sum_{h\in \mathcal{G}} \sope_{h}(t) \Pi_{h}$, and $\Pi_h \in \mathbb{R}^{\vert \mathcal{G} \vert \times \vert \mathcal{G} \vert}$ denotes the unique permutation matrix such that, for any $\pope \in \mathbb{R}^{\vert \mathcal{G} \vert}$ and $\qope = \Pi_h \pope$, we have $\pope_g = \qope_{(hg)}$. For each given sequence $\sope(0),\sope(1),...$, equation \eqref{absevo} looks like the transition dynamics of a (time-inhomogeneous) {\it Markov chain} on the distribution $\pope(t)$ over $\mathcal{G},$ in the sense that the corresponding $\tilde M(t)$ are a sequence of doubly stochastic matrices. In fact, since $(\Pi_h \, \pope)_g = \pope_{h^{-1}g}$, $\tilde M(t)$ implements the (group) convolution of $p(t)$ with $s(t)$.

Definition \ref{def:goal0} is satisfied independently of the particular actions associated to $\mathcal{G}$ if we can ensure convergence to a vector $\pope$ such that:
\begin{equation}
\pope_g  = \pope_{h^{-1}g} \,\,\,\,\,\,\forall\,g,h\,\in\mathcal{G} \, .
\end{equation}
Since for $g$ fixed $\{h^{-1}g : h\in \mathcal{G} \} = \mathcal{G}$, this is consistently equivalent to
\begin{equation}\label{zecond}
\pope_g = 1/\vert\mathcal{G} \vert =: \hat{\pope}_g\,\,\,\forall\,g\,\in\mathcal{G} \, ,
\end{equation}
in accordance with Proposition \ref{prop:3}. To attain symmetrization, we thus require that the dynamics of $\pope$ converges to the unique value $\pope=\hat{\pope}$ given by \eqref{zecond}.

The targeted convergence to a uniform distribution $\hat{\pope}$ under switched dynamics \eqref{absevo} with doubly stochastic transition matrix $\tilde{M}$, is reminiscent of the standard average consensus problem between $\vert\mathcal{G}\vert$ agents in $\R$. There are however at least two major differences between these frameworks.
\begin{enumerate}
	\item The state $\pope(t)$ models ${\cal E}_{t,0}$ from the original problem. In particular, $\pope(0)$ models $\Ei_{0,0}$ which is the identity. Hence, in principle, we would only need to study the evolution from this \emph{known initial state.}
	\item The transition matrix has a different structure inherited from its constituents. For average consensus the transition matrix is essentially the identity plus a sum of symmetric edge-interaction-matrices, with $4$ nonzero entries of equal magnitude per edge of the graph. For $\pope,$ it is a sum of permutation matrices, each of them with $\vert \mathcal{G} \vert$ nonzero entries.
\end{enumerate}
The second point actually alleviates the first one: by group translation, convergence to $\hat{\pope}$ from the particular initial condition $\pope(0)$ corresponding to identity $\Ei_{0,0}$, implies convergence to $\hat{\pope}$ from \emph{any} initial convex weights vector $\pope(0)$. The following section investigates when the system defined by \eqref{absevo} converges to symmetrization. The resemblance with classical consensus will guide us to derive convergence conditions, although they will have to be translated to match the $\pope(t)$ and $\sope(t)$ structure (see second point).


\subsection{Associated continuous-time dynamics}\label{ssec:main2:CT}

A standard procedure to obtain continuous-time dynamics corresponding to the abstract symmetrization framework is to take infinitesimal steps of \eqref{absevo}:
$$\pope(t+dt) = (1-\beta dt) \pope(t) + \beta dt \tilde{M}(t)\pope(t) \;\; $$
and the limit for $dt$ going to zero gives
\begin{equation}\label{eq:CTdyns}
\tfrac{d}{dt} \pope(t) = -\beta \tilde{L} \pope(t) \;\; \text{ with } \;\; \tilde{L} = I_{\vert\mathcal{G}\vert}-\tilde{M} \; ,
\end{equation}
where $\beta > 0$ is just a scalar gain (i.e.~it governs the continuous-time speed). The matrix $\tilde{L}$ in \eqref{eq:CTdyns} is a Laplacian matrix for a balanced graph, as is standard in conventional average consensus, with all off-diagonal elements $\leq 0$, all diagonal elements $\geq 0$, and satisfying $\tilde{L} \hat{\pope} = \tilde{L}^T \hat{\pope} = 0$ i.e.~symmetrization is a stationary solution.

The present paper shall focus on the discrete-time iteration \eqref{absevo}. Similar convergence results for the continuous-time dynamics and discussions for a particular application can be found in our paper \cite{OurMTNS2015}.


\subsection{Example: $\pope(t)$ for gossip consensus}\label{ssec:exmain2}
Let us quickly formulate the gossip algorithm in the action-independent form. In Section \ref{ssec:exmain1}, we illustrated how $x(t+1) = A(t) x(t)$, with
\begin{eqnarray*}
A(t) & = & (1-\alpha)\,I_{mn} + \alpha(t)\, P_{(j,k)}
\end{eqnarray*}
when edge $(j,k)$ is selected at time $t$. The doubly-stochastic $\tilde{M}(t)=(1-\alpha)I+\alpha \Pi_{(j,k)}$ describing the $\pope(t)$ dynamics has dimensions $m! \times m!$ (independently of $n$), with two nonzero entries \emph{on each row and column}: $\tilde{M}_{g,g} = (1-\alpha)$ and $\tilde{M}_{g,\pi_{(j,k)}g} = \alpha$ for all $g\in \mathcal{G}$.
The corresponding continuous-time dynamics would have as nonzero entries $\tilde{L}_{g,g}=\alpha$ and $\tilde{L}_{g,\pi_{(j,k)}g} = -\alpha$ for all $g\in \mathcal{G}$, when the link $(j,k)$ is active.

Convergence of the $\pope$-dynamics is not necessary for convergence of the linear gossip algorithm. Indeed, a dimension counting argument suffices to show that the corresponding actions of $\Perm$ are not linearly independent for $m\geq 4$: the space of possible actions has dimension $m^2$ (consider $A(t) = I_n \otimes A_m(t)$ and count the number of entries in matrix $A_m(t)$), while there are $m!$ permutations and $m! > m^2$ for $m\geq 4$. This means that ensuring convergence of the switched $\tilde{M}$ dynamics for $\pope$ is in principle more demanding than for the switched $A$ for $x$. However, as we prove in the next section, convergence on $\pope$ follows from the typical assumptions of consensus, and allows us to draw conclusions that are valid for all possible $\X$ and actions of $\Perm$.


\section{Convergence analysis}\label{sec:convergence} 
We now examine the convergence properties of \eqref{absevo} with a switching signal $\sope(t)$. This reduces to analyzing an infinite product of doubly stochastic matrices $\tilde{M}(t)$. This problem has been investigated in much detail in other contexts, including standard linear consensus~\cite{TsitsiklisThesis,JADBABAIE,olfati04,Moreau2005}. Among others, \cite{olfati04} proposes a common quadratic Lyapunov function for all possible switchings, which shows that instability is not possible. The question is then, under which conditions is $\hat{\pope}$ \emph{asymptotically} stable. We first give convergence results for deterministic $\sope(t)$. Their adaptation to a randomly selected $\sope(t)$ is explained at the end of the section.


\subsection{Formal conditions and convergence proof}\label{ssec:stab1}
In the context of consensus on graphs, a sufficient condition for convergence is given in terms of a requirement that the union of all edges that appear during a uniformly bounded time interval, must form a connected graph at all times (see e.g.~\cite{Moreau2005}). This result could be applied to \eqref{absevo}, if we view each group element as a node of a \emph{Cayley graph} and draw the directed edges that correspond to the group translations $\Pi_g$ with $\sope_g(t)\geq\underline{\alpha}>0$ at time $t$. The problem at hand however has more structure: an arbitrary adjacency matrix for a graph on $N$ nodes has order $N^2$ parameters, while \eqref{absevo} shows that $\tilde{M}(t)$ is defined by $m!=N$ elements only --- namely the vector $\sope(t)$. In fact we can define a vector of convex weights $\qope_g(t,T)$ such that the evolution from time $t$ to time $t+T$ writes
\begin{equation}
\label{endevo2}
\prod_{i=0}^{T-1}\tilde{M}(t\!+\!i) = \sum_{g\in \mathcal{G}}\qope_g(t,T)\, \Pi_{g} \, .
\end{equation}
This again involves only $m!=N$ elements $\qope_g(t,T)$.
We therefore give independent convergence proofs, in the hope to highlight the role of the assumptions in a way that is more natural in the group-theoretic framework. We next formulate a condition that essentially translates the connected-graph requirement (in fact rather its essential consequence, i.e.~that the transition matrix from $t$ to $t+T$ is primitive) into our framework.
\begin{assump}\label{assum}
Assume the sequence $\sope(t)$ to be such that there exist some finite $T,\delta>0$, such that for each time $t$:
\begin{equation}\label{hyp1}
\qope_g(t,T) > \delta \quad \forall g \in \mathcal{G} \, .
\end{equation}
\end{assump}
This assumption can be translated into properties of the transition matrices in \eqref{absevo}. If $M(t)=M$ for each $t$, then the assumption is equivalent to $M$ being primitive. In the general case, we request that each $\prod_{i=0}^{T-1}\tilde{M}(t\!+\!i)$ is primitive, with all entries at least $\delta.$ 

Notice how Assumption \ref{assum} does not require that $\{ g\in \mathcal{G} : \sope_g(i) > \delta \text{ for some } i \in [t,t+T] \} = \mathcal{G}\, .$ Thus a priori, the (combination of) available actions for all $t$ may be restricted to a subset $\mathcal{S}$ of $\mathcal{G}$; a necessary condition for Assumption \ref{assum} to hold is then that $\mathcal{S}$ generates $\mathcal{G}$. This is similar to requiring that the union of edges appearing during a time interval $T$ in the corresponding Cayley graph form a connected graph, but not necessarily the complete graph. We will further examine Assumption \ref{assum} in Section \ref{ssec:stab2}. 

Now let us formally establish that Assumption \ref{assum} is a sufficient condition to ensure convergence to $\hat \pope$. 

\begin{theorem}\label{thm:conv}
For any switching sequence $\sope(t)$ satisfying Assumption \ref{assum}, the algorithm \eqref{absevo} makes any initial condition $\pope(0)$ converge to the uniform vector $\hat \pope$, elementwise with exponential convergence factor $(1-\vert \mathcal{G}\vert \delta)^{1/T}$.
Furthermore, the Euclidean norm $||\pope-\hat{\pope}||^2$ is a Lyapunov function. 
\end{theorem}

\proof
We can uniformly bound the evolution of the entries of $\pope(N\cdot T)$ for integers $N$ and show that they converge to $1/\vert \mathcal{G} \vert$ at the announced rate.

Consider the sequences of numbers $y(k)$ and $x(k)$ given by:
\begin{eqnarray}
y(k+1) & = & (1-|\mathcal{G}|\delta)y(k) + \delta \quad \mbox{with}\quad y(0)=0,\\
x(k+1) & = & (1-|\mathcal{G}|\delta)x(k) + \delta \quad \mbox{with}\quad x(0)=1,
\end{eqnarray} 
or equivalently since $0<\delta \leq 1/|\mathcal{G}|$ (the minimal entry of $\pope$ cannot be larger than for the uniform distribution),
\begin{equation}\nonumber
y(k) = \tfrac{1}{\vert \mathcal{G}\vert} - \tfrac{1}{\vert \mathcal{G}\vert}(1-|\mathcal{G}|\delta)^k \; , \quad x(k) = \tfrac{1}{\vert \mathcal{G}\vert} + \tfrac{\vert \mathcal{G}\vert-1}{\vert \mathcal{G}\vert}(1-|\mathcal{G}|\delta)^k \, . 
\end{equation}
These two sequences respectively increase / decrease monotonously and exponentially towards $1/|\mathcal{G}|$. Hence we conclude the first part of the proof by showing that
\begin{equation}
\label{ineq01}
x(k)\geq \pope_g(k\cdot T)\geq y(k)
\end{equation}
for every integer $k$ and every $g$. We do this by induction on $k$.

For $k=0$ we have of course $x(0)=1 \geq \pope_g(0) \geq y(0)=0$. Now assuming that the inequality holds for $k$, let us prove that it then holds for $k+1$. For each $g \in \mathcal{G}$, we have
\begin{eqnarray*}
\pope_g((k+1)T) & = & \sum_h \qope_h(t,T) \pope_{h^{-1} g}(kT)
= \delta \sum_h \pope_{h^{-1} g}(kT) + \sum_h (\qope_h(t,T)-\delta) \, \pope_{h^{-1} g}(kT) \\
& = & \delta + \sum_h (\qope_h(t,T)-\delta) \, \pope_{h^{-1} g}(kT)
\end{eqnarray*}
since $\sum_h \pope_{h^{-1} g}(kT)=1$ for each $g$. From the assumptions $\pope_{h^{-1} g}(kT) \geq y(k)$ and $\qope_h(t,T)>\delta$, and using $\sum_h \qope_h(t,t')=1$   for all $t,t'$, we then get:
\begin{eqnarray*}
\pope_g((k+1)T) & \geq & \delta + \sum_h (\qope_h(t,T)-\delta) \, y(kT) \geq \delta +(1-|\mathcal{G}|\delta)y(k)=y(k+1) \, .
\end{eqnarray*}
An analog reasoning shows that $\pope_g((k+1)T) \leq x(k+1)$. 

The exponential convergence of the Euclidean norm for $t$ being a multiple of $T$ is a direct consequence of the exponential elementwise convergence. The fact that for \emph{any} admissible switching sequence this Lyapunov function never increases between \emph{any $t$ and $t+1$}, is shown as follows. Denoting $\dagger$ the transpose of a vector or matrix and $I$ an identity matrix of appropriate dimension, we have
\begin{eqnarray*}
\Vert \pope(t+1) - \hat{\pope} \Vert^2 & = & (\pope(t+1) -\hat{\pope})^\dagger (\pope(t+1) -\hat{\pope}) \\
& = & (\tilde{M}(t)\pope(t) -\hat{\pope})^\dagger (\tilde{M}(t)\pope(t) -\hat{\pope}) \\
& = & \Vert \pope(t) - \hat{\pope} \Vert^2 + \pope(t)^\dagger (\tilde{M}(t)^\dagger \tilde{M}(t)-I) \pope(t) \, .
\end{eqnarray*}
by using $\tilde{M}(t) \hat{\pope} = \hat{\pope}$. 

\noindent Since $\tilde{M}(t)^\dagger \tilde{M}(t)$ is doubly stochastic and symmetric, $(\tilde{M}(t)^\dagger \tilde{M}(t)-I)$ is negative semidefinite for any $t$.\qed

We observe (see appendix) that the relative entropy, or Kullback-Leibler pseudo-distance \cite{cover-thomas} between $\pope(t)$ and $\hat{\pope}$ can also be used as a Lyapunov function to show asymptotic convergence, although in that case it is not as direct to show that convergence is exponential.

As an immediate corollary, we have symmetrization on $\X$ with the associated actions, for {\em any $\X$, any linear group action and any $\sope(t)$ satisfying Assumption \ref{assum}}.
\begin{cor}
Any algorithm of the form \eqref{eq:xevol} on a vector space $\X$ with $\sope(t)$ satisfying Assumption \ref{assum}, asymptotically converges to
$\lim_{t\rightarrow +\infty} x(t)=\bar{\Fi}(x(0)).$ The convergence is exponential and at least as fast as  $(1-\vert \mathcal{G}\vert \delta)^{t/T}.$
\end{cor}

If the actions associated to group elements are linearly dependent, as is the case for consensus, a faster convergence speed can be expected, since convergence at the group level, for the lifted $\pope$ dynamics, is not necessary for convergence of the state.


\subsection{Examining switching signals}\label{ssec:stab2}
Let us now provide some typical examples of switching signals $\sope(t)$ and check if they satisfy Assumption~\ref{assum}. It is actually instructive to start by listing some cases that lead to a violation of the assumption. \begin{itemize}
	\item If (possibly after some initial transient) the vector $\sope(t)$ contains a single nonzero entry at any time, then $\qope(t,T)$ will also contain a single element.
	\item Consider that (after some initial transient) $\sope_g(t)$ can be nonzero at any time only for $g\in\mathcal{S},$ a \emph{subgroup} of $\mathcal{G}$. Then each $\tilde{M}(t)$ is a weighted sum of $\Pi_g$ with $g\in\mathcal{S}$, and by subgroup properties the propagator $\prod_{i=0}^{t-1} \, \tilde{M}(i)$ is also a weighted sum of $\Pi_g$ with $g$ restricted to $\mathcal{S}$, such that we can have $q_g(t,T)\neq 0$ for at most all $g\in \mathcal{S}$.
	\item More generally, if $\sope_g(t)$ can be nonzero at any time only for $g\in\mathcal{S},$ now being some subset of $\mathcal{G}$, and the elements of $\mathcal{S}$ do not generate the whole group, then Assumption \ref{assum} cannot hold.
	\end{itemize}
Conversely, sufficient conditions for Assumption \ref{assum} to hold include the following.
\begin{itemize}
	\item If there exists a set $\mathcal{J} \subset \mathcal{G}$ that generates $\mathcal{G}$ and such that for each $t$, there exists $i \in [t,t+T]$ such that $\mathcal{S}_i = \{ g \in \mathcal{G} : \sope_g(i) > \delta \}$ contains $\mathcal{J} \cup \{ \mathrm{e}_{\mathcal{G}} \}$, then Assumption \ref{assum} is satisfied. We leave this simple proof to the reader.
	\item If $\mathcal{G}$ is Abelian, then the order in which the group elements are selected has no importance, but it is still relevant to know which ones are selected at the same time or not. Then we can use a reduced Cayley graph to investigate Assumption \ref{assum} as follows. For each time $t$, take the set $\mathcal{S}_t = \{ g \in \mathcal{G} : \sope_g(t) > \delta \}$, choose one $\bar{g}_t \in \mathcal{S}_t$ and let $\bar{\mathcal{S}}(t) = \{\bar{g}_t^{-1} g :  g \in \mathcal{S}\setminus\{\bar{g}_t\}\, \}$. Then consider a starting time $t_0$ and recursively construct a graph as follows. Start with a single node $\mathrm{e}_{\mathcal{G}}$. At each step $i=1,2,...,T$, add edges (and potentially vertices) to connect every vertex $h\in \mathcal{G}$ that is already present in the graph at step $i-1$, with the set of nodes $\{ s \, h : s \in \bar{\mathcal{S}}_{t_0+i} \}$. If for all $t$ we have $\sope_{\mathrm{e}_{\mathcal{G}}}(t) > \delta$, and for all $t_0$ the graph obtained at $i=T$ contains all the $g\in\mathcal{G}$, then Assumption \ref{assum} is satisfied.
\end{itemize}

\subsection{Randomized Convergence}
So far we have always formulated convergence properties for a given switching signal $\sope(t)$. We now briefly indicate how they can be adapted when $\sope(t)$ is selected at random. We thus consider that at each time $t$, $\sope(t)$ is selected from a set $\mathfrak{S}$ according to some given probability distribution, independently of the $\sope(i)$ for $i\neq t$. In other words, the $\sope(t)$ are independent, not necessarily identically distributed, random variables over a set of vectors of convex weights. Then we get the following convergence result.
\begin{theorem}\label{thm:stochconv}
Assume that there exist some fixed values of $T$,$\delta$, and $\varepsilon>0$ for which the statement of Assumption \ref{assum} holds with probability at least $\varepsilon$ at each time $t$. Then for any $\gamma>0$, the probability of having an Euclidean distance $\Vert \pope(t)-\hat\pope \Vert <\gamma$ converges to $1$ as $t$ converges to $+\infty$.
\end{theorem}

\proof Assume that Assumption \ref{assum} holds for all times between $t_0$ and $t_0+N_\gamma T$ for some $N_\gamma > 0$. Then we can apply Theorem \ref{thm:conv} between $t_0$ and $t_0+N_\gamma T$, and the resulting exponential convergence is guaranteed to reach $\Vert \pope(t_0+N_\gamma T)-\hat\pope \Vert <\gamma$ for $N_\gamma$ sufficiently large. (Note that the exponential convergence proof of Theorem \ref{thm:conv}, in particular the bounding by sequences \label{sequence},\label{sequence1}, holds for any $\pope(t_0)$.) Moreover, as proved at the end of Theorem \ref{thm:conv}, the Lyapunov function $\Vert \pope(t)-\hat\pope \Vert$ cannot increase between $t$ and $t+1$ under \eqref{absevo}, for any vector of convex weights $\sope(t)$. Hence we would also have $\Vert \pope(t)-\hat\pope \Vert <\gamma$ for any $t > t_0+N_\gamma T$.

The proof is concluded by noting that, under the specified random choice of the signal $\sope(t)$, the probability that a sequence of $B\cdot N_\gamma\cdot T$ elements contains no subsequence of $N_\gamma T$ consecutive elements satisfying Assumption \ref{assum}, is at most $(1-\varepsilon^N)^B$. The latter converges to $0$ as $B$ goes to $\infty$, thus as $t$ goes to $\infty$ for fixed $\gamma$,$\delta$,$T$. 

\qed

Let us briefly discuss some examples of randomized evolutions.\begin{itemize}
\item If at each time, we randomly select a single element $h(t)$ from $\mathcal{G}$ with probability of $h(t)=g$ being greater than zero for all $g,$ and take 
\begin{equation}\label{eq:gtype}
\sope_{h(t)}(t)=\alpha \;, \quad \sope_{\mathrm{e}_{\mathcal{G}}}(t)=(1-\alpha)\;, \quad \sope_g(t) = 0 \text{ for } g\notin \{ h(t),\,\mathrm{e}_{\mathcal{G}} \} \, ,
\end{equation}
then the requirements of Theorem \ref{thm:stochconv} are clearly satisfied. Of course this situation directly generalizes to cases where more than one $h(t) \in \mathcal{G}$ is applied at each time.
\item Like in the deterministic case, a similar result is obtained if in \eqref{eq:gtype} we randomly select $h(t)$ from some subset $\mathcal{S}$ of $\mathcal{G}$, and this subset generates the whole group. The subset may also vary (e.g.~cyclically) with time, as long as it allows with nonzero probability to construct one sequence satisfying Assumption \ref{assum}. The linear gossip algorithm fits in this category, as the connected graph condition in Proposition \ref{prop:00} ensures that swaps of adjacent agents can be selected in a way that generates the whole group of permutations.
\end{itemize}

\noindent A few remarks are in order.
\begin{remark}[Time-varying possibilities] {\em Theorem \ref{thm:stochconv} only requires some uniform upper bound $T$ on a time interval that guarantees that all group elements are associated with weights of at least $\delta>0$. It thus allows for dynamics where $\pope(t)$ does not evolve towards $\hat\pope$ for shorter time intervals, as long as there is a nonzero probability to reduce the distance from $\hat\pope$ in finite time.  Therefore, we can ensure convergence if, for example, one strictly contractive evolution is applied only every $T_0$ steps, while we do not know how $\sope_g$ is selected in between. }
\end{remark}
\begin{remark}[Explicit robustness to $\alpha$] {\em A major contribution of Theorem \ref{thm:stochconv} is to establish the \emph{robustness} of consensus-like algorithms with respect to uncertainties in the values of $\sope_g(t)$ for a wide variety of applications (see Section~\ref{sec:apps}). Indeed, if we consider that the $h \in \mathcal{S}$ for which $\sope_h \neq 0$ are chosen deterministically, but the values $\sope_h(t)$ are randomly chosen in some compact set strictly inside $[0,1]$ for all $t$, then Assumption \ref{assum} holds with given $T$ either for all such sequences or for none; in the former case, compactness ensures that $\delta$ is bounded from below, and Theorem \ref{thm:stochconv} holds. This shows that it is not important to control the exact proportions in which the chosen actions are applied. Typically in a gossip algorithm \cite{BoydGossip}, one uses the maximally mixing value $\alpha=1/2$. Nonetheless, convergence holds provided that $\alpha(t) \in [\underline{\alpha},\overline{\alpha}] \subset (0,1)$ for all $t$. Of course, the choice of $\sope(t)$ can severely affect convergence \emph{speed}, but this discussion goes beyond the scope of the present paper.}
\end{remark}
\begin{remark}
{\em In relation with Assumption \ref{assum}, it is useful to work with sequences satisfying (with a given non-zero probability) $\sope_{\mathrm{e}_{\mathcal{G}}}(t) \geq \beta$ at any $t$ for some constant $\beta>0$. Indeed, this ensures that once $q_g(t,t+t_1)\geq \delta' > 0$ for some $t_1\leq T$, we have $q_g(t,t+T) \geq \delta = \delta' \beta^{T-t_1}$. Most results in linear consensus~\cite{TsitsiklisThesis,ConsensusReview,Moreau2005} explicitly make this assumption. Not assuming $\sope_{\mathrm{e}_{\mathcal{G}}}(t) \geq \beta > 0$ for all $t$ generally makes it necessary to perform a detailed analysis of the successions in $\sope(t)$ in order to ensure Assumption \ref{assum}.}
\end{remark}


\section{Examples}\label{sec:apps} We next illustrate the potential of our results by illustrating a variety of tasks covered by our framework. For these tasks, the gossip-inspired dynamics we have studied recover some relevant, existing class of algorithms or variations of these. We naturally start with consensus-type problems, including in Example \ref{ssec:app:quantum} a quantum consensus algorithm which we have proposed and analyzed with a rather technical, ad-hoc approach in \cite{OurJournal1}. With the new {\em lifted} convergence results at hand, the solution is immediate. We then turn to more general symmetrization problems which do not include a network structure or a consensus-type task. These include random state generation protocols and quantum dynamical decoupling, two key tasks in quantum information theory and applications. In order to further illustrate the variety of the potential applications, we also include an academic example, showing how even the seemingly unrelated discrete Fourier transform can be seen as a symmetrization problem. The analysis of these protocols from a unified symmetrization viewpoint, and hence explicit proof of their robustness and randomization properties, are, to the best of our knowledge, new results. The list of examples is by no means assumed to be exhaustive, and we are confident that more areas of application will be identified.


\subsection{Linear consensus}\label{ssec:app:ccons}
The gossip algorithm of Section \ref{sec:gossip} is one basic application of our framework. The group-theoretic language also encompasses other basic linear algorithms for average consensus of $m$ subsystems in $\R^n$.

The most standard consensus algorithm implements, at each time, a motion of each subsystem towards the average of its neighbors in an \emph{undirected graph} $G(t)$. Thus the edges of $G(t)$ model a set of interactions that are all simultaneously active. This corresponds to setting $\sope_g(t)\neq 0$ for $g=e$ and for all $g\in\Perm$ that model a pairwise permutation of two agents linked by an edge in $G(t)$, up to possibly having to use negative 
$\sope_g(t)$. We recall that, since the actions associated to $\Perm$ in standard consensus are not linearly independent, this is not the only way to lift the consensus dynamics to the permutation group; in particular, there is a way to do this without ever necessitating negative $\sope_g(t)$, see next paragraph. Gossip, with a single edge active at a time and hence only two nonzero elements in $\sope_g(t)$, is just a particular case. 

In the group-theoretic formulation, there seems no reason to limit our algorithmic building blocks to pairwise permutations. Including more general permutations allows one to cover situations with explicit multipartite interactions, {\em e.g.}~where subsystem $1$ forwards its value to $2$, who simultaneously transmits its value to $3$, and so on. Selecting $\sope_g \neq 0$ specifically for $g$ corresponding to such situations, allows to model \emph{synchronous} linear consensus iterations with symmetric or non-symmetric state transition matrix $A(t)$. The resulting $A(t)$ however will still be doubly-stochastic for any $\sope$.
As proved by Birkhoff \cite{birkhoff-stoch}, any doubly stochastic matrix can be decomposed as a convex sum of permutations. 
The corresponding network structure is called a \emph{balanced directed graph}~\cite{olfati04}, and one could argue that the interpretation as a sum of general permutations gives a sensible rationale as why a graph might be ensured to be balanced in the consensus context. In this sense, any consensus algorithm on a balanced directed graph can be seen as a generalization of a gossip-type algorithm. Convergence, independently of the particular application, is guaranteed if Assumption \ref{assum} is satisfied.\\

Let us consider a concrete example of a consensus application: three vehicles need to establish agreement about the position of the center of a circle, on which they will move as a formation \cite{RodA}. Let $x_k \in \mathbb{R}^2$ denote the center estimate for vehicle $k$, with $k=1,2,3$. We assume that vehicles $2$ and $3$ cannot communicate. This corresponds to a consensus problem for a graph on 3 nodes $\{1,2,3\}$ and with edges $(1,2)$, $(1,3)$. A compatible consensus algorithm is:
\begin{eqnarray}
\nonumber x_1(t+1) & = & (1-2\alpha) x_1(t) + \alpha x_2(t) + \alpha x_3(t) \\
\label{eq:exampleConvSp} x(t+1) = A x(t) \quad : \quad x_2(t+1) & = & (1-\alpha) x_2(t) + \alpha x_1(t) \\
\nonumber x_3(t+1) & = & (1-\alpha) x_3(t) + \alpha x_1(t) \;
\end{eqnarray}
with $\alpha \leq 0.5$ to maintain double stochasticity.

From the symmetrization viewpoint, this problem considers all possible permutations of the initial estimates of the circle centers associated to the 3 vehicles: 
\begin{equation}\label{exchrow}
\begin{array}{lcccccc}
\text{permutation} & x_1(0) &        x_1(0) &            x_2(0) &           x_3(0) &           x_2(0) &          x_3(0) \\
                 & x_2(0) &        x_3(0) &            x_1(0) &           x_2(0) &           x_3(0) &          x_1(0) \\
                 & x_3(0) &        x_2(0) &            x_3(0) &           x_1(0) &           x_1(0) &          x_2(0) \\
\text{weight}& {\pope_e}  & {\pope_{[1,3,2]}} & {\pope_{[2,1,3]}} & {\pope_{[3,2,1]}} & {\pope_{[2,3,1]}} &  {\pope_{[3,1,2]}}\\
\end{array}
\end{equation}
The vector $\pope(t)$ represents the weight distribution over these 6 situations, labeling each permutation $\pi$ of $[1,2,3]$ with the vector $[\pi(1),\pi(2),\pi(3)]$. 
According to \eqref{eq:totevol}, at any time $x_1(t)$ is the sum of the first element of each of the 6 columns, weighted by the corresponding entry of $\pope(t)$. One can similarly compute $x_2(t)$ and $x_3(t)$.
We start with all the weight concentrated on the trivial permutation, corresponding to $\pope_e(0) = 1$. The consensus dynamics redistributes the weight such that finally all six situations have the same weight i.e.~$\pope = \hat{\pope}$, the vector with all elements equal to $1/6$. When $\pope = \hat{\pope}$, the average positions of $x_1$, $x_2$ and $x_3$ are all the same and located at the barycenter of $x_1(0)$, $x_2(0)$ and $x_3(0)$, as expected from average consensus.

Following \eqref{absevo}, the lifted dynamics associated to \eqref{eq:exampleConvSp} would be modeled by:
\begin{equation}\label{eq:NoForITS}
\sope_e = 1-2\alpha \; ; \quad \sope_{[2,1,3]} = \alpha \; ; \quad \sope_{[3,2,1]} = \alpha \; ; \quad \sope_g = 0 \text{ for all other } g \, .
\end{equation}
For example, the action associated to $[2,1,3]$, corresponding to active communication along the link $(1,2)$, can be viewed as exchanging the first and second row of \eqref{exchrow}. Equivalently, leaving the first three rows of \eqref{exchrow} in place, the action associated to $[2,1,3]$ ``exchanges weight'' between $\pope_{e}$ and $\pope_{[2,1,3]}$, between $\pope_{[3,2,1]}$ and $\pope_{[2,3,1]}$, and between $\pope_{[3,1,2]}$ and $\pope_{[1,3,2]}$.\\

We have mentioned that convergence in the permutation group is not necessary for convergence of the corresponding consensus algorithm. Related to this point, convergence \emph{speed} may differ for $\pope$ and for $x$. This can be illustrated already on the above simple example.
The eigenvalues of the $\tilde{M}$ matrix corresponding to \eqref{eq:NoForITS} indeed differ from those of the $A$ matrix associated to consensus in \eqref{eq:exampleConvSp}. For $\alpha > 0.4$ we get $\sigma(\tilde{M}) > \sigma(A)$, where $\sigma(X)$ denotes the dominating singular value of $X$ i.e.~the largest modulus among all eigenvalues of $X$ that differ from $1$. Thus for $0.5 \geq \alpha > 0.4$, the eigenvalues of $\tilde{M}$ which govern convergence on the permutation group, underestimate the actual convergence speed of \eqref{eq:exampleConvSp} on $\mathbb{R}^6$. For instance $\alpha = 0.45$ gives a geometric convergence rate with factor $\sigma(A)=0.55$ for consensus, but only with $\sigma(\tilde{M})=0.8$ on the permutation group. Intuitively this can be understood by noting that the circle centers on the above schematic representation would all be located at the same central position already if e.g.~$\pope_e = \pope_{[3,1,2]} = \pope_{[2,3,1]} = 1/3$. Hence converging to $\pope = \hat{\pope},$  while it is actually attained by the algorithm \eqref{eq:exampleConvSp}, is not necessary for reaching consensus towards controlling the circular formation. Therefore the effective convergence speed can be faster for the original, ``un-lifted'' dynamics.


\subsection{Gossip symmetrizing probability distributions}\label{ssec:app:proba}
Consider a collection of $m$ subsystems, each one possessing a random variable $y_j$ on the same outcome set $Y$, for $j=1,2,\ldots,m$. We denote $\mathbb{P}$ the joint probability distribution of the $y_j$. In order to maintain a compact notation we will consider $Y$ countable, but the uncountable case does not present additional technical difficulties. We are interested in symmetrizing the joint probability distribution, i.e.~attaining a distribution $\hat{\mathbb{P}}$ such that
\begin{eqnarray}
\label{eq:probsym}  & & \hat{\mathbb{P}} [y_1=a_1,...,y_j=a_j,...,y_k=a_k,...,y_m=a_m] \\
\nonumber			& & = \hat{\mathbb{P}} [y_1=a_1,...,y_j=a_k,...,y_k=a_j,...,y_m=a_m]
\end{eqnarray}
for all choices of $j,k$ and of the considered outcomes $\{ a_i \}$. The invariance then also holds for general permutations in $\Perm$. We want to achieve this in a distributed way, where at each time $t$ a reduced set $E(t)$ of pairwise interactions are available.

Our framework suggests the following randomized way to perform this task. At each time $t$ a pair $(j,k)$ is selected from $E(t)$, the random variables at these locations are swapped with probability $\alpha$, and remain in place with probability $1-\alpha$. This random action still leaves $y_j(t+1),y_k(t+1)$ two random variables on $Y$, but their probability distributions have changed: e.g.~the new random variable $y_j(t+1)$ at location $j$ follows the marginal distribution of $y_j(t)$ with probability $1-\alpha$, or it follows the marginal distribution of $y_k(t)$, with probability $\alpha$. Overall, {\em not knowing whether the random variables have been exchanged or not}, the resulting probability distribution for the $y_i(t+1)$, $i=1,2,...,m$ writes:
\begin{eqnarray}
\lefteqn{\mathbb{P}_{t+1}[y_1=a_1,...,y_j=a_j,...,y_k=a_k,...,y_m=a_m] \; =} \phantom{KKKK} \label{eq:probalgo}\\
& &(1-\alpha) \; \mathbb{P}_{t}[y_1=a_1,...,y_j=a_j,...,y_k=a_k,...,y_m=a_m] \nonumber\\
& & + \alpha \; \mathbb{P}_{t}[y_1=a_1,...,y_j=a_k,...,y_k=a_j,...,y_m=a_m] \nonumber
\end{eqnarray}

In the group symmetrization picture, this framework (goal \eqref{eq:probsym} and dynamics \eqref{eq:probalgo}) corresponds to the exact same setting as standard gossip consensus, with $\mathcal{G}=\Perm$ the group of permutations on $m$ objects. Only the action is different, now implementing a swap on probability distributions (\emph{including all correlations with other random variables} than the ones involved in the swap), instead of a swap of real numbers.


\subsection{Gossip symmetrizing quantum subsystems}\label{ssec:app:quantum} A classical random variable can be viewed as a special, commutative case in the framework of quantum, non-commutative probability theory. Following this analogy, the previous example can be extended to quantum observables -- that is, self-adjoint linear operators on some Hilbert space $\Hi$. This is done in \cite{OurJournal1} with an ad-hoc approach, independently of the present general framework.

Consider a multipartite quantum system, composed of $m$ isomorphic subsystems with individual Hilbert space $\Hi_1 = \Hi_2 = ... = \Hi_m$. The state of the overall system, which has the role of a probability distribution, is described by a density operator $\rho$ on the tensor product of the individual Hilbert spaces, $\Hi=\Hi_1\otimes\Hi_2\otimes...\otimes\Hi_m$.  Let $\X$ be the set of self-adjoint operators on $\Hi,$ associated to observable physical quantities. With $\mathcal{G}$ still being the permutation group of $m$ objects, represented on the integers $1,2,...,m$ by elements $\pi$, we define the action $a_q(\pi,X)$ on $\X$ by
$$a_q(\pi,X) = X_{\pi(1)} \otimes X_{\pi(2)} \otimes ... \otimes X_{\pi(m)}$$
for operators of the form $X = X_1 \otimes X_2 \otimes ... X_m$ on $\Hi$, and extend it to the whole set $\X$ of self-adjoint operators on $\Hi$ by linearity. To each such action, we can associate a unitary operator $U_\pi$ on $\Hi$ such that
$$a_q(\pi,X) = U_\pi^\dagger \, X \, U_\pi \quad \text{for all } X \in \X\, ,$$
where $U^\dagger$ denotes the adjoint of $U$ (i.e.~the complex conjugate transpose in matrix notation).

For this quantum system, the group dynamics corresponding to linear gossip would apply at each step a convex combination of the identity and the permutation of two physical subsystems $j,k$. Explicitly, the dynamics of $X$ is given by:
$$
X(t+1)=(1-\alpha) X(t) +\alpha U_{(j,k)}^\dagger X(t) U_{(j,k)},\quad \alpha\in[0,1] \, .
$$ 
This is a completely-positive, trace-preserving and unital map on $\X$. The latter two properties mirror double stochasticity of $\tilde M(t)$.

The convergence of the action-independent dynamics to $\hat \pope$ directly implies that both the cyclic and randomized versions of this quantum gossip algorithm will drive any initial $X \in \X$ to
$$\hat X =\frac{1}{m!}\;{\textstyle\sum_{\pi\in\mathfrak{P}}}\; U_\pi^\dagger\, X\, U_\pi\; .$$
Physically, this implies that the measurement of any joint property on a subset of $n<m$ quantum systems will give the same statistics irrespective of the particular $n$ subsytems that are selected.

Equivalently, we could consider as $\X$ the set of all density operators on $\Hi$, with the action $a'_q(g,\cdot) := a_q(g^{-1},\cdot)$. These two equivalent viewpoints on quantum mechanics are well-known as the ``Heisenberg picture'' and the ``Schr\"odinger picture''. Example \ref{ssec:app:proba} is retrieved when all considered operators are diagonal in a fixed basis, and the diagonal of the density operator is then equivalent to a classical probability density. In the language of \cite{OurJournal1}, this dynamics attains \emph{symmetric state consensus}.


\subsection{Randomized discrete Fourier transform}\label{ssec:app:Fourier}

The above applications all involve permutations as the underlying group. The permutation group and the set of generators that can be activated encodes the network structure for the distributed computation task. We next show, starting with an academic example, how {\em the same class of algorithms can be used to tackle different problems that do not involve any network or consensus-reaching task. Specifically, a choosing a different group structure can lead to a \emph{randomized} algorithm computing the discrete Fourier transform.}

The discrete Fourier transform of a (column) vector $x=(x_0,x_1,...,x_{N-1}) \in \mathbb{C}^N$ is the (column) vector $\chi = (\chi_0,\chi_1,...,\chi_{N-1})$ with
\begin{equation}\label{eq:F1}
\chi_k = \frac{1}{N} \sum_{n=0}^{N-1} e^{-i\,\frac{k\,n\, 2\pi}{N}}\, x_n \quad \text{ for } k=0,1,...,N-1	\, ,
\end{equation}
up to normalization\footnote{Our developments can be extended to functions on finite Abelian groups, with the Fourier transform defined on characters.}.
The complex numbers $\{e^{i\,k\, 2\pi/N} : k=0,1,...,N-1 \}$ characterizing the Fourier transform form a faithful representation of the cyclic group of order $N$, that is the Abelian group generated by a single element $\bar{g}$,
$$\mathcal{G}_{c,N}=\{\, \mathrm{e}=\bar{g}^0=\bar{g}^N, \bar{g},\bar{g}^2,\bar{g}^3,...,\bar{g}^{N-1} \,\} \, .$$
We next show how the computation of \eqref{eq:F1} can be obtained as a byproduct of a symmetrization task with respect to an action of $\mathcal{G}_{c,N}$.

It is convenient to consider the vector space $\mathbb{R}^{N \times N}$ and associate to the (column) vector $x \in \mathbb{R}^N$ the square matrix $\,X = x\, \mathbf{1}^T \,$, where $\mathbf{1}^T$ is the row vector of ones. To $\bar{g} \simeq e^{i\, 2\pi/N}$ we associate the group action $a(\bar{g},\cdot)=Q(\cdot)$ defined by:
\begin{eqnarray}\label{F-action}
	X \; \mapsto &\;\; &  Q(X) = \sigma \, X \, D^{-1} \\[2mm]
	\nonumber \text{with} & & D = \text{diag}(1,\,e^{i\, 2\pi/N},\,e^{i\, 4\pi/N},...,e^{i\, (N-1)2\pi/N}) \\
	\nonumber & & \sigma = \left({\scriptsize  \begin{array}{cccccc} 
	0 & 1 & 0 & 0 & ... & 0 \\
	0 & 0 & 1 & 0 & ... & 0 \\
	0 & 0 & 0 & 1 & ... & 0 \\
	\vdots \\
	0 & 0 & 0 & 0 & ... & 1 \\ 
	1 & 0 & 0 & 0 & ... & 0  \end{array}} \right) \, .
\end{eqnarray}
The action corresponding to a general group element is obtained by composition. Direct computation shows that the $m,n$ element of $\hat X = \frac{1}{N} \sum_{k=0}^{N-1} \, Q^k(X)$, resulting from the symmetrization of $X$ under the action $Q$, equals
\[ \hat{X}_{[m,n]} \; = \; \frac{1}{N}\sum_{k=0}^N\; x_{(m+k\hspace{-2mm}\mod (N-1))}\; e^{-i\frac{2\pi k}{N}n} \,.\]
Hence symmetrization under this action of $\mathcal{G}_{c,N}$ gives the Fourier transform of $x$ as:
\[ \chi^T=
\left[
\begin{array}{ccccc}
  1 & 0 & 0 & \ldots & 0   
\end{array}
\right] \, \hat{X} \, .
\]
The robust convergence of algorithm \eqref{absevo} thus indicates that the Fourier transform does not necessarily have to be computed in an orderly fashion, but can asymptotically result from rather arbitrary convex combinations of the actions $Q^k$ with different $k$, as long as the $\sope(t)$ ensure sufficient mixing. Note that the actions $\{Q^0,Q^1,...,Q^{N-1}\}$ are all linearly independent, so the map from dynamics on group actions to dynamics on $\pope$ is one-to-one.


\subsection{Random state generation}\label{ssec:app:rand}
A variety of applications require to generate random numbers, codewords or, more generally, {\em states} with a target probability distribution. This includes among others the Markov chain Montecarlo methods~\cite{MC01} as well as classical and quantum cryptography protocols \cite{nielsen-chuang}. A typical, and fundamental, target probability distribution is the uniform or Haar measure on compact sets.
Random sample generators must hence be able to transform some \emph{generic} source of randomness -- i.e.~not necessarily uniform nor in fact exactly known -- into a (almost) \emph{uniform} probability distribution. There are various ways of doing this, and our framework points to a particular class of so-called random circuits~\cite{QRC-Science2003,QRC-Lloyd2005}. Indeed, group symmetrization provides a robust way to obtain a uniform distribution on a finite set of states $\Y$ that are linked by a group of transformations ${\cal G}$, if we can pick elements of ${\cal G}$ with some generic probability distribution.

More precisely, consider a finite group ${\cal G},$ and its linear action $a(g,\cdot)$ on a vector space $\X.$
For some fixed $y_{\rm e} \in \X$, consider its {\itshape orbit}, i.e. the set   $\mbox{Orb}_{{\cal G}}(y_{\rm e})=\{y_g=a(g,y_{\rm e}),\,g\in{\cal G}\}$. We want to generate a state $y(T)$ that is uniformly (pseudo-)randomly distributed over $\mbox{Orb}_{{\cal G}}(y_{\rm e})$, by passing a deterministic $y(0) \in \mbox{Orb}_{{\cal G}}(y_{\rm e})$ through a sequence of (pseudo-)random operations, labeled for convenience by time $t=0,1,...,T-1$. Each operation is associated to a  $g(t) \in \mathcal{G}$, drawn according to some possibly unknown probability distributions $\sope_g(t),$ mutually independent at each time. We make the technical assumption that $g\neq h \Rightarrow a(g,y(0)) \neq a(h,y(0))$ i.e.~$|\mbox{Orb}_{{\cal G}}(y_{\rm e})|=|{\cal G}|$.

As $y$ propagates through the sequence according to $y(t+1) = a(g(t),y(t))$, the probability $\pope_h(t)$ to have $y(t+1) = a(h,y(0))$ follows dynamics \eqref{absevo}. Hence according to  Theorem \ref{thm:conv}, it is sufficient that $\sope(t)$ allows to satisfy Assumption \ref{assum} to ensure that the distribution of $y(T)$ converges to the \emph{uniform} distribution over $\mbox{Orb}_{{\cal G}}(y_{\rm e})$ as $T\rightarrow\infty$. Note that for a fixed circuit distribution $\sope_g(t)$, we indeed apply Theorem \ref{thm:conv} as we are modeling the \emph{deterministic} evolution (as $t$ increases) of a probability distribution.

\begin{remark}
In addition to finite groups, the case in which $\mathcal{G}$ becomes a continuous Lie group is of great interest for practical applications, including quantum information and more specifically random quantum circuit theory \cite{QRC-Science2003,QRC-Lloyd2005}. In that framework, the space of interest is associated to a register of $N$ quantum bits, so that $\X \cong \mathbb{C}^{2^N}$; the group of physically relevant unitary evolutions for the register, or {\em gates}, is $\mathcal{G} = SU(2^N)$. The finite group setting can effectively approximate such continuous distribution by considering a sufficiently dense subset of the Lie group. It is well known \cite{nielsen-chuang} that there exist finite {\em universal sets} of gates which generate a mathematically dense subset of $SU(2^N)$; ensuring $\sope_g(t) > 0$ on such a universal set, is sufficient to satisfy Assumption \ref{assum} for any finite subset of a dense subset of $SU(2^N)$.
\end{remark}


\subsection{Dynamical decoupling}\label{ssec:app:DD}
{\em Quantum Dynamical Decoupling} (DD) is a set of open-loop control techniques that are primarily used to reduce the effect of unknown Hamiltonian drifts, or couplings to the environment, on a target quantum system~\cite{viola-DD}. The main idea is to apply a sequence of ``switching'' unitary rotations to the system, such that effects of the undesired dynamics over the sequence of unitary rotations compensate each other and the net effect is negligible.
This task can be translated into a symmetrization task \cite{zanardi-symmetrizing}, and we show here how our results suggest a robust DD scheme. For the sake of simplicity, we restrict ourselves to the suppression of the drift Hamiltonian in finite dimensional systems. The extension to decoupling from the environment is straightforward.

The quantum evolution of an isolated finite-dimensional system is driven by its Hamiltonian $H$, a Hermitian matrix whose spectrum is associated to the energy levels of the system.
The propagator for the system is then the unitary operator
\[U_t=e^{-iH t}\]
when $H$ is constant. When $H$ is time-varying, the propagator must be computed as an ordered product of exponentials over infinitesimal intervals. The resulting unitary operator can be associated to an effective Hamiltonian $H_{eff}$ such that 
$$U_{T} = e^{-iH_{\text{eff}}\,T} \; .$$
A DD strategy consists in a time-dependent control Hamitlonian $H_c(t)$ such that, for any constant $H_d$ in a class of expected perturbations, the effective Hamiltonian associated to $H_d+H_c(t)$ is ``close'' to a scalar matrix after a predefined time $T$: $H_{\text{eff}} \approx \lambda I$ with $\lambda \in \mathbb{R}$. Indeed, this would suppress any physical effect of $H_d$ at time $T$ since global phases of the form $U_t=e^{i\lambda t}$ are irrelevant for predictions in quantum mechanics \cite{sakurai}. DD in its simplest form entails a sequence of fast, impulsive control operations that induce a group of ``instantaneous'' unitary transformations on the system, and achieves first-order suppression of $H_d$. The relevant time interval $[0,T)$ is subdivided into $N$ subintervals of length $dt=T/N$ and instantaneous controls are applied at the end of each sub-interval so that the effective Hamiltonian for subinterval $[(k-1)\, dt,\; k\,dt)$ is $g_k H_d g_k^\dagger$ with $g_k \in \mathcal{G}$. Then, the Magnus expansion~\cite{magnus} allows to approximate the exact evolution from time $0$ to $T$ to first order as:
\begin{equation}\label{eq:DDframes}
e^{-i \, dt\, g_1 H_d g_1^{\dagger}} \; e^{-i \, dt\, g_2 H_d g_2^{\dagger}}\; ... \; e^{-i \, dt\, g_N H_d g_N^\dagger} \;\;\approx\;\;
e^{-i\, dt \; {\textstyle \sum_{k=1}^{N}} \, g_k H_d g^{\dagger}_k} \; =: e^{-i\,T\,\bar{H}}\; ,
\end{equation}
where $\vphantom{k}^{\dagger}$ denotes matrix conjugate transpose. Accuracy improves as the product of $H_d$ with $dt$ gets smaller.
Hence, given a class $\mathfrak{H}_0$ of drift Hamiltonians on some finite-dimensional Hilbert space $\Hi \cong \mathbb{C}^n,$ first-order DD follows from identifying a finite subgroup $\mathcal{G}$ of unitaries such that
\begin{equation}\label{eq:Haway}
\tfrac{1}{| \mathcal{G} |} \; \sum_{g \in \mathcal{G}} \; g \, H_d \, g^{\dagger} = \lambda I
\end{equation}
for all $H_d \in \mathfrak{H}_0$. In the language of our paper, DD achieves symmetrization with respect to a group $\mathcal{G}$, and the latter is selected  such that the action $a(g,H) = g\, H\, g^{\dagger}$ on the space $\X$ of all Hamiltonians $H$ satisfies $\bar{\mathcal{F}}(\mathfrak{H}_0) \subseteq \{ \lambda I, \lambda \in \mathbb{R} \}$.

Achieving symmetrization in \eqref{eq:DDframes} means choosing each $g \in \mathcal{G}$ an equal number of times over the $N$ subintervals. An obvious choice is just to take $N=m|\mathcal{G}|$ and iterate $m$ times a predefined path through the elements of $\mathcal{G}$. However, when $H_d$ is not really constant for a duration $|\mathcal{G}|\, dt$ or when considering higher-order Magnus terms, the potential advantage of randomized~\cite{viola-randomizedDD,santos-randomizedDD06} or concatenated~\cite{Khodjasteh2005} sequences of $g_k$ has been recognized. Our general dynamics \eqref{absevo} allows to retrieve and combine these two variants of DD and, in particular, to highlight their robustness.

Consider an iterative construction of the sequence of unitaries $g_k$, where at the $n$-th iteration the time interval $[0,T)$ is subdivided into $N=2^n$ subintervals. Denote ${\cal S} \subseteq {\cal G}$ the set of available control actions. We start at $n=0$ from the situation with no control pulses, so $g_1= \mathrm{e} \cong I_{\Hi}$  over $[0,T)$ and $\bar H=H_d$. Increasing $n$, we then choose one element $h(n) \in {\cal S}$, we divide each subinterval $\left[(m-1)\tfrac{T}{2^n},\, m\tfrac{T}{2^n}\right)$ into two equal time intervals 
$\left[(2m-2)\tfrac{T}{2^{n+1}},\, (2m-1)\tfrac{T}{2^{n+1}}\right)$ and $\left[(2m-1)\tfrac{T}{2^{n+1}},\, 2m\tfrac{T}{2^{n+1}}\right)$, and we update the sequence as follows for $m=1,...,2^n$:
{\begin{equation}\label{eq:ddalg}
\text{At } n: \;\; g_m  = \bar{g} \quad\;\; \Rightarrow \quad\;\; \text{At } n+1: \;\; g_{2m-1} = \bar{g} \,, \;\; g_{2m} = h(n)\bar{g} \, .
\end{equation}
Denoting by $\pope_g(n)$ the fraction of time $[0,T)$ during which $g_k = g \in \mathcal{G}$,} the procedure \eqref{eq:ddalg} correponds to \eqref{absevo} with $t$ replaced by $n$, and the switching signal:  
\begin{equation}\label{eq:ddsope}
 \sope_g(n) = 1/2 \; \text{ for } g \in \{ \mathrm{e}_{\mathcal{G}}, \, h(n) \}\; , \quad  \sope_g(n) = 0 \; \text{ for all other } g \in \mathcal{G}\, .
\end{equation}
In action form, the average Hamiltonian at the $n$-th iteration is
$$ \bar{H}_n= {\textstyle \sum_{g \in \mathcal{G}}} \; \pope_g(n) \, a(g,H_d) \, = \, {\textstyle \sum_{g \in \mathcal{G}}} \; \sope_g(n-1) \, a(g,\bar{H}_{n-1}) \, .$$
Our theorems ensure the convergence of $\bar{H}_n$ towards the $\mathcal{G}$-symmetrized form \eqref{eq:Haway} of $H_d$ as $n$ is increased, if Assumption \ref{assum} holds. This is valid both for deterministic or random choices of the $h(n)$. Furthermore, our results indicate a remarkable generality and robustness of the procedure: (i) the control actions $h(n)$ don't have to be chosen uniformly in $\mathcal{G}$, actually any deterministic choice or probabilistic distribution over enough elements will work; (ii) the set $\mathcal{S}$ of control actions does not have to be all $\mathcal{G}$, e.g.~a set of generators would be sufficient; and (iii) the subdivision can be more general than a ``perfect average'': any $\sope_{h(n)}(n) = 1-\sope_{\mathrm{e}}(n) = \alpha$ with $\alpha \in (0,1)$ would asymptotically work, not just \eqref{eq:ddsope} where $\alpha=1/2$.


\section{Conclusion}\label{sec:conclusion}
The present paper shows how the simple dynamics of linear gossip consensus can inspire robust iterative procedures for tasks that can be formulated as \emph{symmetrization with respect to a finite group.} We prove convergence for a general symmetrization process with either deterministic or randomized choices of the individual iterations. We have shown how a variety of existing algorithms, some unrelated to any network structure, are covered by the framework. We expect that in many other applications the \emph{robustness} of the consensus formulation can be advantageously carried over to symmetrizations tasks, e.g.~including actions on infinite-dimensional spaces. Natural directions for expanding our results in the short term include the development of (approximate) symmetrization procedures for infinite and continuous groups, as well as an in-depth study of convergence \emph{speed} for specific protocols. Regarding the latter, our bound in Theorem \ref{thm:conv} can be unnecessarily pessimistic especially when the concerned group actions are not linearly independent, as is the case e.g.~for consensus. The possibility to lift, to the abstract symmetrization framework, several speed-up strategies for faster mixing is also being investigated. Replacing the linear action on a vector field by abstract algebraic structures could also offer a rewarding way to unify more algorithmic procedures, hopefully including e.g.~alternating directions optimization or dominant eigenvector computations, under the symmetrization viewpoint.

\section{Acknowledgments}
The authors would like to thank Lorenza Viola for suggesting the application of these techniques to random quantum circuits, for pointing out some key references and for numerous, fruitful and pleasant discussions on these topics. This work has been partially supported by the QUINTET and QFUTURE strategic projects of the Dept. of Information Engineering and University of Padua, and by the Belgian Inter-University Attraction Poles network DYSCO.

\bibliographystyle{plain}
\bibliography{bibToSIAM.bib}

\begin{thebibliography}{10}

\bibitem{altafini-tutorial}
C.~Altafini and F.~Ticozzi.
\newblock Modeling and control of quantum systems: An introduction.
\newblock {\em IEEE Trans. Aut. Cont,}, 57(8):1898 --1917, 2012.

\bibitem{awerbuch1987optimal}
Baruch Awerbuch.
\newblock Optimal distributed algorithms for minimum weight spanning tree,
  counting, leader election, and related problems.
\newblock In {\em Proceedings of the nineteenth annual ACM symposium on Theory
  of computing}, pages 230--240. ACM, 1987.

\bibitem{MC01}
B.A. Berg, D.P. Landau, W.S. Kendall, R.~Chen, and E.A. Thompson.
\newblock {\em {M}arkov {C}hain {M}onte {Carlo}: innovations and applications}.
\newblock World Scientific Publishing, 2005.

\bibitem{birkhoff-stoch}
G.~Birkhoff.
\newblock Three observations on linear algebra.
\newblock {\em Univ. Nac. Tucuan. Revista A}, 5:147--151, 1946.

\bibitem{BoydGossip}
S.~Boyd, A.~Ghosh, B.~Prabhakar, and D.~Shah.
\newblock Randomized gossip algorithms.
\newblock {\em IEEE Trans. Information Theory (Special issue)},
  52(6):2508--2530, 2006.

\bibitem{carli2011pi}
Ruggero Carli, Edoardo D'Elia, and Sandro Zampieri.
\newblock A pi controller based on asymmetric gossip communications for clocks
  synchronization in wireless sensors networks.
\newblock In {\em Decision and Control and European Control Conference
  (CDC-ECC), 2011 50th IEEE Conference on}, pages 7512--7517. IEEE, 2011.

\bibitem{cover-thomas}
T.~M. Cover and J.~A. Thomas.
\newblock {\em Elements of Information Theory}.
\newblock Wiley, 1991.

\bibitem{dimakis2010gossip}
Alexandros~G Dimakis, Soummya Kar, Jos{\'e}~MF Moura, Michael~G Rabbat, and
  Anna Scaglione.
\newblock Gossip algorithms for distributed signal processing.
\newblock {\em Proceedings of the IEEE}, 98(11):1847--1864, 2010.

\bibitem{Donoho2006cs}
D.L. Donoho.
\newblock Compressed sensing.
\newblock {\em IEEE Trans.Information Theory}, 52(4):1289--1306, 2006.

\bibitem{Alggrpth}
C.~D. Godsil and Gordon. Royle.
\newblock {\em Algebraic graph theory}.
\newblock Springer New York, 2001.

\bibitem{ishii2010distributed}
Hideaki Ishii and Roberto Tempo.
\newblock Distributed randomized algorithms for the pagerank computation.
\newblock {\em Automatic Control, IEEE Transactions on}, 55(9):1987--2002,
  2010.

\bibitem{JADBABAIE}
A.~Jadbabaie, J.~Lin, and A.S. Morse.
\newblock Coordination of groups of mobile autonomous agents using nearest
  neighbor rules.
\newblock {\em IEEE Trans. Automatic Control}, 48(6):988--1001, 2003.

\bibitem{QRC-Lloyd2005}
J.Emerson, E.Livine, and S.Lloyd.
\newblock Convergence conditions for random quantum circuits.
\newblock {\em Physical Review A}, 72(6):060302, 2005.

\bibitem{QRC-Science2003}
J.Emerson, Y.~Weinstein, M.~Saraceno, S.~Lloyd, and D.~Cory.
\newblock Pseudo-random unitary operators for quantum information processing.
\newblock {\em Science}, 302:2098--2100, 2003.

\bibitem{Khodjasteh2005}
K.Khodjasteh and D.A.Lidar.
\newblock Fault-tolerant quantum dynamical decoupling.
\newblock {\em Phys. Rev. Lett.}, 95(18):180501, 2005.

\bibitem{viola-generalnoise}
E.~Knill, R.~Laflamme, and L.~Viola.
\newblock Theory of quantum error correction for general noise.
\newblock {\em Phys. Rev. Lett.}, 84(11):2525--2528, 2000.

\bibitem{leonard2007collective}
Naomi~Ehrich Leonard, Derek~A Paley, Francois Lekien, Rodolphe Sepulchre,
  David~M Fratantoni, and Russ~E Davis.
\newblock Collective motion, sensor networks, and ocean sampling.
\newblock {\em Proceedings of the IEEE}, 95(1):48--74, 2007.

\bibitem{santos-randomizedDD06}
L.F.Santos and L.Viola.
\newblock Enhanced convergence and robust performance of randomized dynamical
  decoupling.
\newblock {\em Phys. Rev. Lett.}, 97(15):150501, 2006.

\bibitem{magnus}
W.~Magnus.
\newblock On the exponential solution of differential equations for a linear
  operator.
\newblock {\em Commun. Pure and Appl. Math.}, 7:649--673, 1954.

\bibitem{OurJournal1}
L.~Mazzarella, A.~Sarlette, and F.~Ticozzi.
\newblock Consensus for quantum networks: from symmetry to gossip iterations.
\newblock {\em IEEE Trans. Automatic Control}, 60(1):158--172, 2015.

\bibitem{moallemi2006consensus}
Ciamac~Cyrus Moallemi and Benjamin Van~Roy.
\newblock Consensus propagation.
\newblock {\em Information Theory, IEEE Transactions on}, 52(11):4753--4766,
  2006.

\bibitem{Moreau2005}
L.~Moreau.
\newblock Stability of multi-agent systems with time-dependent communication
  links.
\newblock {\em IEEE Trans. Automatic Control}, 50(2):169--182, 2005.

\bibitem{Motwani1995}
R.~Motwani and P.~Raghavan.
\newblock {\em Randomized algorithms}.
\newblock Cambridge University Press, 1995.

\bibitem{ConvOptCons}
A.~Nedic and A.~Ozdaglar.
\newblock Distributed subgradient methods for multiagent optimization.
\newblock {\em Automatic Control, IEEE Transactions on}, 54(1):48--61, 2009.

\bibitem{nielsen-chuang}
M.~A. Nielsen and I.~L. Chuang.
\newblock {\em Quantum Computation and Information}.
\newblock Cambridge University Press, Cambridge, 2002.

\bibitem{ConsensusReview}
R.~Olfati-Saber, J.A. Fax, and R.M. Murray.
\newblock Consensus and cooperation in networked multi-agent systems.
\newblock {\em Proc. IEEE}, 95(1):215--233, 2007.

\bibitem{olfati04}
R.~Olfati-Saber and R.M. Murray.
\newblock Consensus problems in networks of agents with switching topology and
  time delays.
\newblock {\em IEEE Trans. Automatic Control}, 49(9):1520--1533, 2004.

\bibitem{pearl1986fusion}
Judea Pearl.
\newblock Fusion, propagation, and structuring in belief networks.
\newblock {\em Artificial intelligence}, 29(3):241--288, 1986.

\bibitem{sakurai}
J.~J. Sakurai.
\newblock {\em Modern Quantum Mechanics}.
\newblock Addison-Wesley, New York, 1994.

\bibitem{RodA}
Rodolphe Sepulchre, Derek~A Paley, and Naomi~Ehrich Leonard.
\newblock Stabilization of planar collective motion: All-to-all communication.
\newblock {\em IEEE Trans.Aut.Control}, 52(5):811--824, 2007.

\bibitem{OurMTNS2015}
F.~Ticozzi, L.~Mazzarella, and A.~Sarlette.
\newblock Symmetrization for quantum networks: a continuous-time approach.
\newblock {\em Proc. Conf.~on Math. Theory of Networks and Systems (MTNS)},
  2014.

\bibitem{TsitsiklisThesis}
J.N. Tsitsiklis and M.~Athans (advisor).
\newblock Problems in decentralized decision making and computation.
\newblock {\em PhD Thesis, MIT}, 1984.

\bibitem{viola-DD}
L.~Viola, E.~Knill, and S.~Lloyd.
\newblock Dynamical decoupling of open quantum system.
\newblock {\em Phys. Rev. Lett.}, 82(12):2417--2421, 1999.

\bibitem{viola-randomizedDD}
Lorenza Viola and Emanuel Knill.
\newblock Random decoupling schemes for quantum dynamical control and error
  suppression.
\newblock {\em Phys. Rev. Lett.}, 94:060502, 2005.

\bibitem{wan2002distributed}
Peng-Jun Wan, Khaled~M Alzoubi, and Ophir Frieder.
\newblock Distributed construction of connected dominating set in wireless ad
  hoc networks.
\newblock In {\em INFOCOM 2002. Twenty-First annual joint conference of the
  IEEE computer and communications societies. Proceedings. IEEE}, volume~3,
  pages 1597--1604. IEEE, 2002.

\bibitem{xiao2005scheme}
Lin Xiao, Stephen Boyd, and Sanjay Lall.
\newblock A scheme for robust distributed sensor fusion based on average
  consensus.
\newblock In {\em Information Processing in Sensor Networks, 2005. IPSN 2005.
  Fourth International Symposium on}, pages 63--70. IEEE, 2005.

\bibitem{zanardi-symmetrizing}
P.~Zanardi.
\newblock Symmetrizing evolutions.
\newblock {\em Phys. Lett. A}, 258:77, 1999.

\end{thebibliography}


\appendix
\section{Convergence in relative entropy}
We here show that the relative entropy, or Kullback-Leibler pseudo-distance, is also a Lyapunov function for the convergence of $\pope(t)$ to $\hat\pope$ under our symmetrizing dynamics. Before giving the proof, let us recall some basic facts about relative entropy and the log sum inequality.
 
The relative entropy, or Kullback-Leibler pseudo-distance \cite{cover-thomas} of a vector of convex weights $\{\qope_g\}_{g\in\mathcal{G}}$ with respect to another one $\{\pope_g\}_{g\in\mathcal{G}}$ is given by:
\begin{equation}
K(\pope\Vert\mathsf{q})=\sum_{g\in \mathcal{G}} \pope_g \left(\log \pope_g-\log \mathsf{q}_g\right).
\end{equation}
This expression is not symmetric in $\pope,\qope$, but $K(\pope\Vert\qope)\geqslant 0$ and the equality holds if and only if $\pope=\qope$. We shall also use the following~\cite{cover-thomas}.
\begin{proposition}[Log Sum Inequality]\label{prop:logsum}
Let $\{a_i\}_{i=1}^n$ and $\{b_i\}_{i=1}^n$ be nonnegative numbers. Then it holds:
\begin{equation}
\sum_{i=1}^n a_i \log \frac{a_i}{b_i}\geqslant \left(\sum_{i=1}^n a_i \right)\log \frac{\sum_ia_i}{\sum_i b_i}.
\end{equation}
Furthermore, excluding the singular cases where $\sum_i a_i = 0$ or $\sum_i b_i = 0$, the equality holds if and only if $\frac{a_i}{b_i}=\alpha$ is constant over $i=1,\dots,n$.
\end{proposition}
\vspace{3mm}

We can now turn to the convergence proof using $K(\pope(t)\Vert\hat{\pope})$ as Lyapunov function. The corresponding statement would be equivalent to Theorem \ref{thm:conv} except that we do not prove the exponential character of the convergence. $K(\pope(t)\Vert\hat{\pope})$ is nonnegative and it equals zero if and only if $\pope(t)=\hat{\pope}$. To use it as a strict Lyapunov function, it remains to prove that, under Assumption \ref{assum}, this relative entropy of $\pope(t)$ with respect to $\hat\pope$ strictly decreases after (any) $T$ steps. For every $t$ we have that:
\begin{equation*}
 \begin{aligned}
&K(\pope(t+T)\Vert\hat{\pope})= \sum_{g\in\mathcal{G}}\pope_g(t+T) \log \frac{\pope_g(t+T)}{\hat{\pope}_g }\\
&\phantom{KK}= \;\; \sum_{g\in\mathcal{G}}\left(\sum_{h\in \mathcal{G}}\qope_h(t,T)\pope_{h^{-1}g}(t) \right)\log \frac{\sum_{h }\qope_h(t,T)\pope_{h^{-1}g}(t)}{\sum_{h }\qope_h(t,T) \hat{\pope}_g }.
\end{aligned}
\end{equation*}
Now by applying the log sum inequality over $h$ for each fixed $g$ we get:
\begin{equation}
\label{eq:fromlogsum}
 \begin{aligned}
& \left(\sum_{h\in \mathcal{G}}\qope_h(t,T)\pope_{h^{-1}g}(t) \right)\log \frac{\sum_{h }\qope_h(t,T)\pope_{h^{-1}g}(t)}{\sum_{h }\qope_h(t,T) \hat{\pope}_g }\\
&\phantom{KK}\leq \;\; \sum_{h\in \mathcal{G}}\; \left( \qope_h(t,T)\pope_{h^{-1}g}(t) \log \frac{ \qope_h(t,T)\pope_{h^{-1}g}(t)}{ \qope_h(t,T) \hat{\pope}_{h^{-1}g}} \right) \,.
 \end{aligned}
\end{equation}
Furthermore, Assumption \ref{assum} allows us: (i) to divide by $\qope_h(t,T)$; and (ii) in conjunction with the fact that $\sum_g \pope_g(t) = 1$ for all $t,$ to exclude the singular cases in Proposition \ref{prop:logsum}. Therefore the equality in \eqref{eq:fromlogsum} holds if and only if
\begin{equation*}
\frac{\qope_h(t,T)\pope_{h^{-1}g}(t)}{\qope_h(t,T) \hat{\pope}_{h^{-1}g}} =  \frac{\pope_{h^{-1}g}(t)}{\hat{\pope}_{h^{-1}g}} 
\end{equation*}
is constant over all $g' = h^{-1}g \in \mathcal{G}$. Since $\sum_{g' \in \mathcal{G}}\pope_{g' \in \mathcal{G}}(t)=\sum_{g'}\hat{\pope}_{g'}=1$ for every $t$, the equality holds if and only if $\pope(t)=\hat{\pope}$.
Returning to the sum over $g$, we thus get
\begin{equation}
0 \leqslant K(\pope(t+T)\Vert\hat{\pope})\leqslant K(\pope(t)\Vert\hat{\pope})
\end{equation}
and each equality holds if and only if $\pope(t)=\hat{\pope}$. Henceforth the Lyapunov function  $K(\pope(t)\Vert\hat{\pope})$ strictly decreases after any $T$ steps, as the requirement $q_h(t,T)>\delta$ ensures that for any given $\pope(t)\neq\hat{\pope}$, we get in \eqref{eq:fromlogsum} a strict contraction factor independent of $\sope(t)$. This ensures, by Lyapunov arguments, that the system asymptotically converges to $\pope=\hat{\pope}$.\vspace{3mm}

The fact that exponential convergence is not as direct, would also require another approach for the randomized case, that is Theorem \ref{thm:stochconv}.

\end{document}